\documentclass[aps,prc,preprint,superscriptaddress,showpacs,showkeys]{revtex4-1}

\usepackage{graphicx}

\begin{document}


\title{Components of the dilepton continuum in Pb+Pb collisions at 
$\sqrt{s_{_{NN}}} = 2.76 $ TeV }
\author{\large V. Kumar}
\author{\large P. Shukla}
\email{pshukla@barc.gov.in}
\affiliation{Nuclear Physics Division, Bhabha Atomic Research Center, Mumbai, India}
\affiliation{Homi Bhabha National Institute, Anushakti Nagar, Mumbai, India}
\author{\large R. Vogt}
\affiliation{Physics Division, 
Lawrence Livermore National Laboratory, Livermore, CA 94551, USA}
\affiliation{Physics Department, University of California, Davis, 
CA 95616, USA}

\date{\today}

\begin{abstract}
The dilepton invariant mass spectrum measured in heavy-ion collisions 
includes contributions from important QGP probes such as thermal radiation and
the quarkonium ($J/\psi$, $\psi'$ and $\Upsilon$) states. Dileptons coming 
from hard $q \overline q$ scattering, the Drell-Yan process, contribute 
in all mass regions. In heavy-ion colliders, such as the LHC, semileptonic 
decays of heavy flavor hadrons provide a substantial contribution to the 
dilepton continuum.  Because the dilepton continuum can provide 
quantitative information on heavy quark yields and their medium modifications, 
it is important to identify which dilepton sources populate different parts of 
the continuum.  In the present study, we calculate $c \overline c$ and 
$b \overline b$ production and determine their contributions to the dilepton
continuum in Pb+Pb collisions at $\sqrt{s_{_{NN}}} = 2.76$ TeV with and without 
including heavy quark energy loss.  We also 
calculate the rates for Drell-Yan and thermal dilepton production.
The contributions to the continuum from these dilepton sources are
studied in the kinematic ranges relevant for the LHC detectors.
The relatively high $p_T$ cutoff for single leptons excludes most dileptons
produced by the thermal medium.  Heavy flavors are the dominant source of
dilepton production in all the kinematic regimes except at forward rapidities 
where Drell-Yan dileptons become dominant for masses greater than 10 GeV/$c^2$.
\end{abstract}



\maketitle

\section{Introduction}

Heavy-ion collisions study the interaction of matter at the extreme 
temperatures and densities where a Quark-Gluon Plasma (QGP), a phase of nuclear
matter dominated by color degrees of freedom, is expected to form.  Experimental
efforts in this field began with the CERN SPS ($\sqrt{s_{_{NN}}} \sim 16-19$~GeV)
and evolved with data \cite{INTRO} from the first heavy-ion collider, the 
Relativistic Heavy-Ion Collider (RHIC) at Brookhaven National Laboratory 
($\sqrt{s_{_{NN}}} = 200$ GeV) in the last decade.  The advent of Pb+Pb 
collisions at $\sqrt{s_{_{NN}}} = 2.76$ TeV at the LHC has increased excitement 
in this field.  One of the most striking QGP signals is quarkonium suppression 
\cite{SATZ}. Quarkonia are identified by their reconstructed mass peaks in the 
dilepton invariant mass distribution.  Below $\sim 12$~GeV/$c^2$, the dilepton
distribution includes a number of resonance peaks: $\rho$, $\omega $ and $\phi$
at low masses and the $\psi$ and $\Upsilon$ states at higher masses.  At
91~GeV/$c^2$, the $Z^0 \rightarrow l^+ l^-$ peak appears.  The continuum beneath
these resonances is primarily composed of leptons from semileptonic decays of
heavy flavor hadrons.  These heavy flavor decays not only contribute to the
resonance background but are important physics signals in their own right
\cite{LastCallLHC,GAVIN,LIN,LIN2,SHUR,DKS}. The continuum yields in 
Pb+Pb collisions compared to those in $pp$ collisions can provide information 
about the medium properties.  This makes it important to know the various 
contributions to the dilepton continuum in different kinematic regimes. 
 
 The first measurements of the dilepton spectra at the LHC have recently been 
reported \cite{zboson,CMSQ,ATLASJ}. The CMS experiment reported the first 
measurements of the $Z^0$ mass region in Pb+Pb collisions \cite{zboson} as
well as measurements of the full dimuon distribution, including quarkonia 
\cite{CMSQ}.  ATLAS has also reported $J/\psi$ and $Z^0$  measurements in the
dimuon channel \cite{ATLASJ}.  The second LHC Pb+Pb run, 
at much higher luminosity, has provided higher statistics measurements of the
dilepton spectra over the full available phase space.  With the measurement of
dilepton spectrum in Pb+Pb collisions at the LHC, it is time to re-examine the 
continuum contributions to the dilepton mass spectrum.  The production cross 
sections of $c \overline c$ and $b \overline b$ pairs at 
$\sqrt{s_{_{NN}}} = 2.76$ TeV are calculated to next-to-leading order (NLO) 
and their correlated contributions to the dilepton continuum are subsequently 
obtained. We also include the effect of energy
loss of charm and bottom quarks in the medium consistent with measurements of 
the suppression factor $R_{AA}$ on the lepton spectra from semileptonic
decays of charm and bottom \cite{RAARHIC,RAAAlice}. These contributions are
compared to direct dilepton production from the Drell-Yan process and from
thermal production in the medium.  We then evaluate the relative importance of
these contributions in the LHC detector acceptances.

While there have been previous studies of Pb+Pb collisions at 5.5 TeV 
\cite{GAVIN,LIN2,GALL}, a re-examination is appropriate at the current, lower, 
center of mass energy and with the final detector acceptances.  In addition, 
updated parameterizations of the parton distribution functions as well as
estimates of the effect of energy loss on single particle spectra
and determinations of the initial temperature from the charged particle
multiplicity are now available and should lead to improved predictions. The
experimental dilepton measurements presently concentrate on
resonances.  However, background-subtracted dilepton continuum measurements
should soon be available with good statistics at 2.76 TeV in both $pp$ and 
Pb+Pb collisions which could be used to infer propeties of the medium 
produced in Pb+Pb collisions.

\section{Dilepton production by hard processes}
\label{harddilep}

Dilepton production from semileptonic decays of $D\overline D$ (charm) and 
$B\overline B$ (bottom) meson pairs has been an area of active theoretical 
\cite{GAVIN,LIN,SHUR,FEIN,MUST} and experimental \cite{RHICSingleEl} research. 
The large heavy quark mass allows their production to be calculated in 
perturbative QCD.  We calculate the production cross sections for 
$c\overline c$ and  $b\overline b$ pairs to NLO in pQCD \cite{GAVIN,LIN} using
the CTEQ6M parton densities \cite{CTEQ6}.  The central EPS09 parameter set 
\cite{EPS09} is used to calculate the modifications of the parton densities in 
Pb+Pb collisions.  

We include the theoretical uncertainty bands on charm and bottom production
following the method of Ref.~\cite{CNV}.  We use the same set of parameters
as that of Ref.~\cite{CNV} with the exclusive NLO calculation of Ref.~\cite{MNR}
to obtain the exclusive $Q \overline Q$ pair rates as well as their decays
to dileptons.  We take $m_c = 1.5$~GeV/$c^2$, $\mu_F/m_T = \mu_R/m_T = 1$ and  
$m_b = 4.75$~GeV/$c^2$, $\mu_F/m_T = \mu_R/m_T = 1$ as the central values for
charm and bottom production respectively.  Here $\mu_F$ is the factorization 
scale, $\mu_R$ is the renormalization scale and $m_T = \sqrt{m^2 + p_T^2}$.  
The mass and scale variations are added in quadrature to obtain the uncertainty
bands \cite{CNV}.  

\begin{figure}
\includegraphics[width=0.48\textwidth]{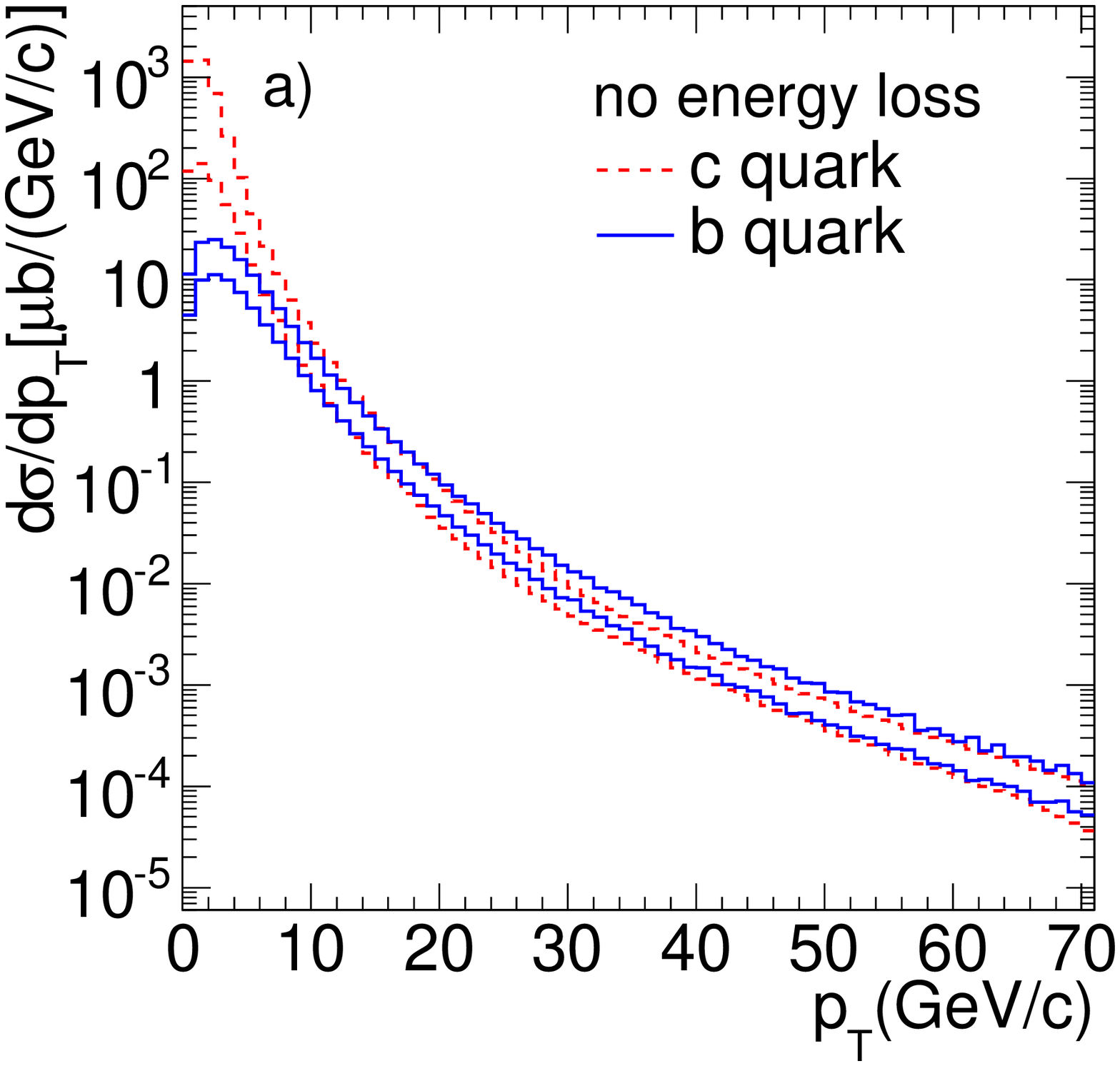}
\includegraphics[width=0.48\textwidth]{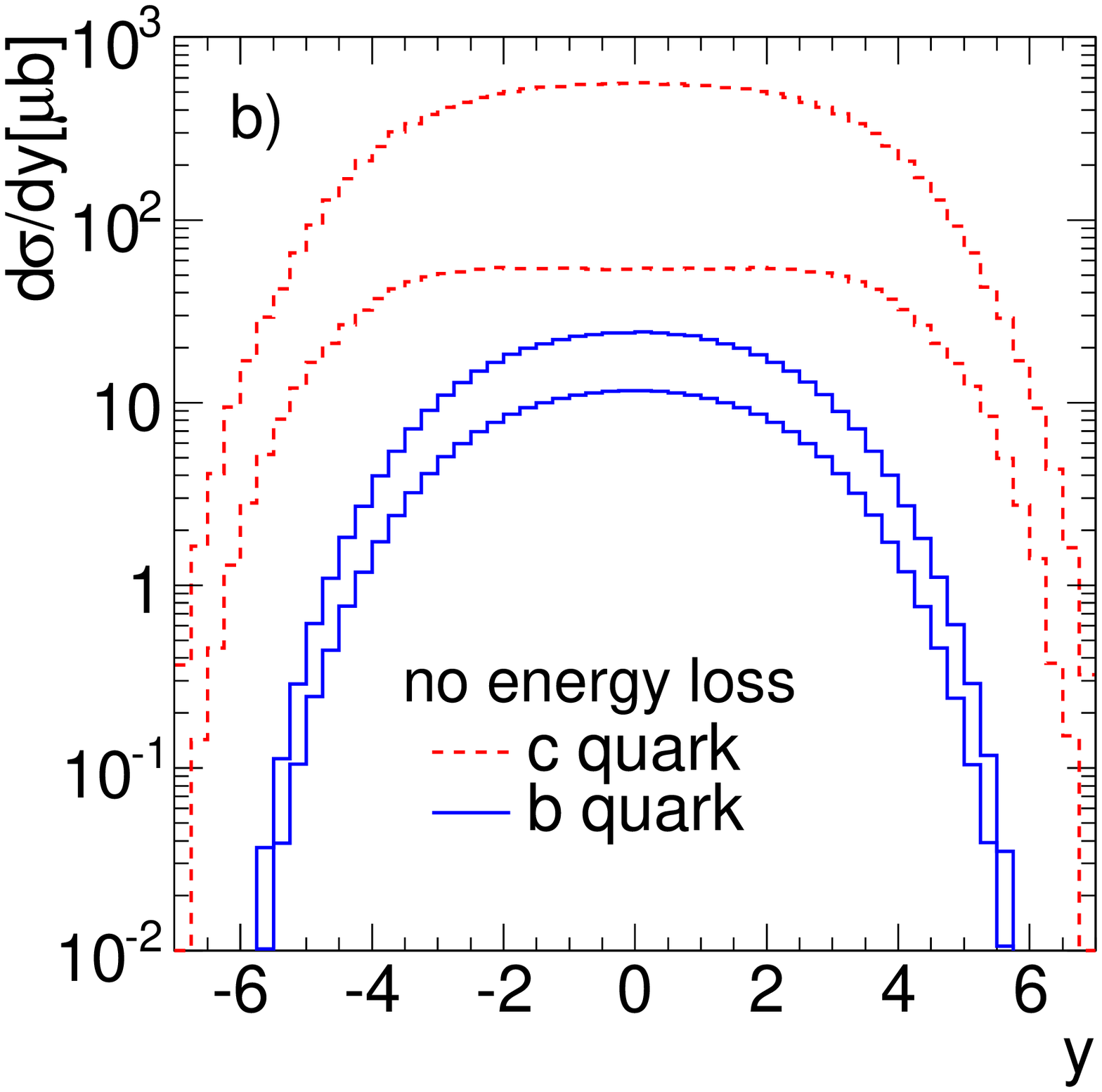}
\caption{(Color online) Theoretical uncertainty bands on inclusive single 
charm and bottom quark production cross sections per nucleon as functions 
of $p_T$ (left) and rapidity (right) for $\sqrt{s_{_{NN}}} = 2.76$ TeV. 
The uncertanities are calculated by 
varying the quark mass, renormalization scale $\mu_{R}$
and factorization scale $\mu_{F}$. The calculations include 
modification of the initial parton distributions with the EPS09 central
parameter set.
No final state energy loss is included.}
\label{QPtYUncertNoLoss}
\end{figure}

Figure~\ref{QPtYUncertNoLoss} shows the uncertainty bands on the $p_T$ 
and rapidity distributions of charm and bottom quarks in Pb+Pb collisions
at $\sqrt{s_{_{NN}}} = 2.76$~TeV with shadowing effects included.  We only
calculate the uncertainties in the production cross sections due to the
mass and scale parameters and not those due to the EPS09 modifications or those
of the parton densities.  Both of these uncertainties are smaller than those
due to the choice of mass and scale \cite{NVF}, particularly for $p_T \geq m$.  
The uncertainties on the heavy flavor production cross sections can be rather
large, see Refs.~\cite{RVjoszo,RVHP08}.  Thus the relative charm and bottom 
rates at 2.76 TeV may vary by a factor of two or more before dense matter
effects such as energy loss are taken into account.  While a recent reevaluation
of the mass and scale parameters used to calculate charm production shows
that the uncertainty on the charm production cross section can be reduced, it 
cannot be eliminated \cite{NVF}.  

The differences in the quark $p_T$ distributions are
primarily at low $p_T$.  For $p_T > 10$~GeV$/c$, the uncertainty bands overlap
almost completely with the upper limit on the bottom production band somewhat
above the charm upper limit for $p_T > 20$~GeV$/c$.
The widths of the rapidity distributions are limited by the heavy quark mass.  
Thus the charm rapidity distribution is broader than that for bottom. 
The uncertainty bands are broader in rapidity than in $p_T$ for charm and the
bands for the two flavors are cleanly separated
because the $p_T$-integrated rapidity distribution is dominated by low $p_T$
where the charm cross section is clearly greater and the scale uncertainties
are larger.

The production cross sections for heavy flavor and Drell-Yan dileptons
at $\sqrt{s_{_{NN}}}= 2.76$ 
TeV are shown in Table~\ref{NLOcros}.  The number of $Q \overline Q$ pairs
in a minimum bias Pb+Pb event is obtained from the per nucleon cross
section, $\sigma_{\rm PbPb}$, by
\begin{eqnarray}
N_{Q \overline Q} = {A^2 \sigma_{\rm PbPb}^{Q \overline Q}  \over  
\sigma_{\rm PbPb}^{\rm tot}} \, \, .
\end{eqnarray}
At 2.76 TeV, the total Pb+Pb cross section, $\sigma_{\rm PbPb}^{\rm tot}$, 
is 7.65 b \cite{PbPbTotal}.

\begin{figure}
\includegraphics[width=0.48\textwidth]{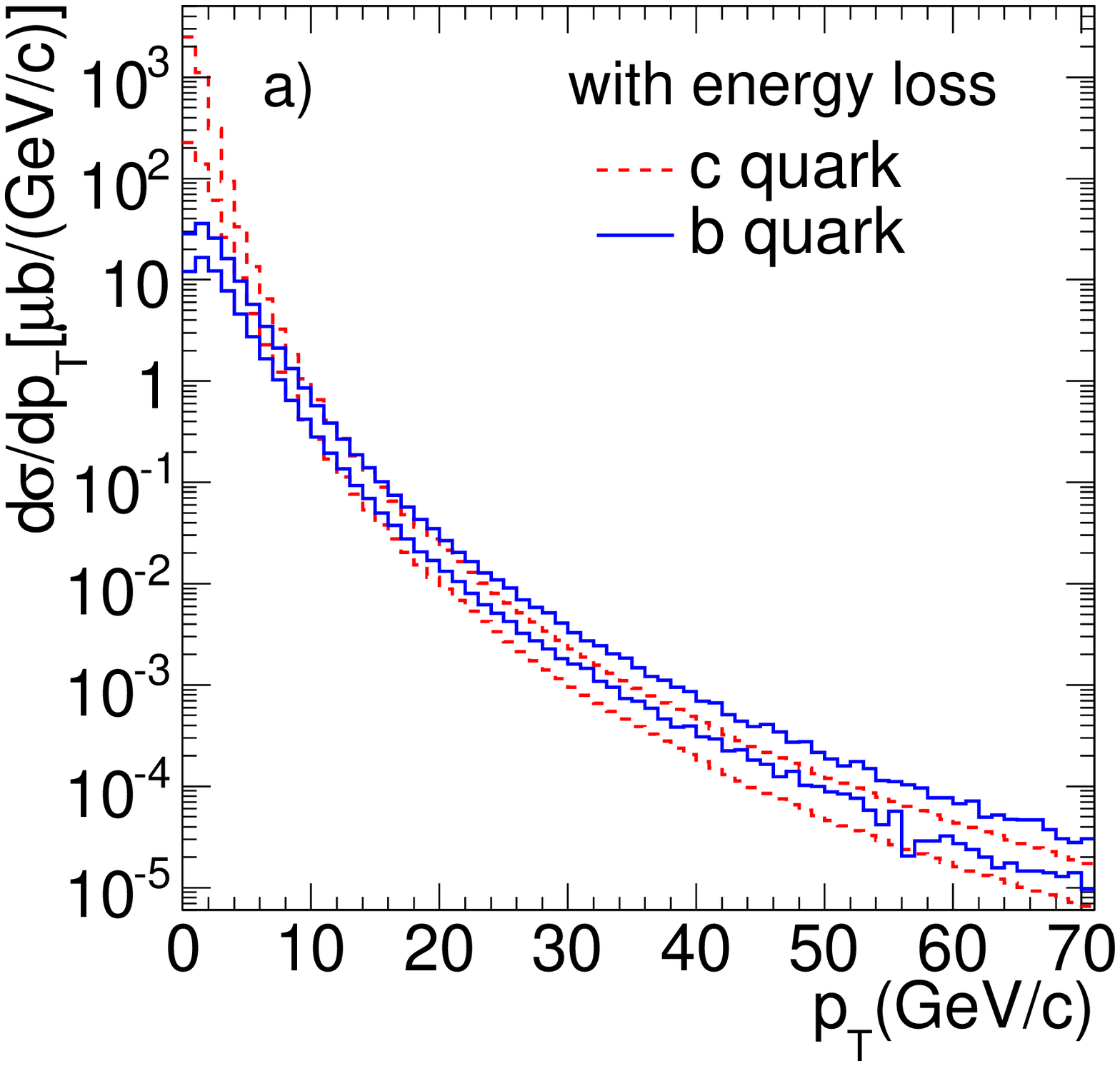}
\includegraphics[width=0.48\textwidth]{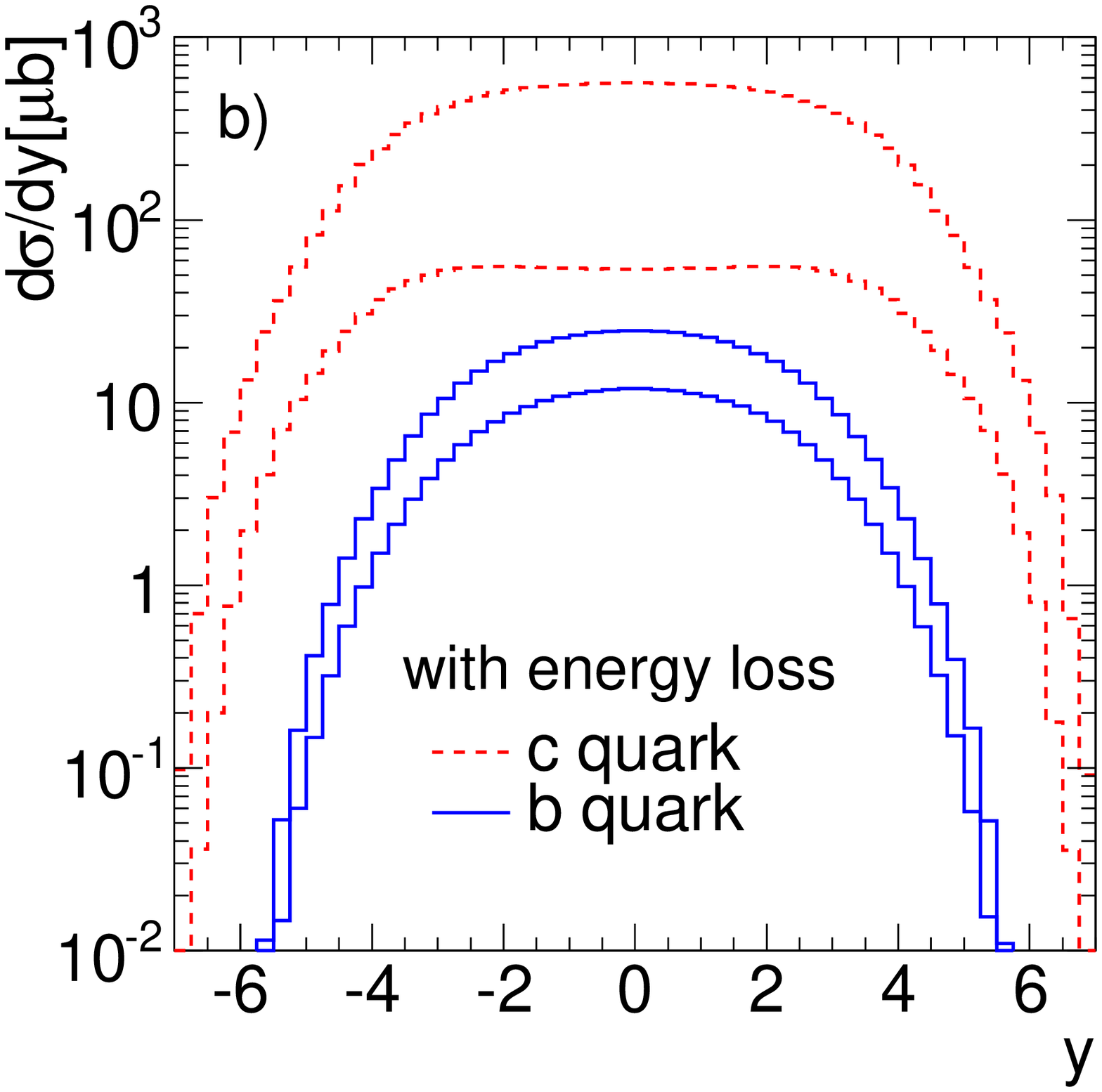}
\caption{(Color online) Theoretical uncertainty bands on inclusive single 
charm and bottom quark production cross sections per nucleon as functions 
of $p_T$ (left) and rapidity (right) for $\sqrt{s_{_{NN}}} = 2.76$ TeV. 
The uncertanities are calculated by 
varying the quark mass, renormalization scale $\mu_{R}$
and factorization scale $\mu_{F}$. The calculations include 
modification of the initial parton distributions with the EPS09 central
parameter set.
Here we include final state energy loss assuming that the charm and bottom
quark $R_{AA}$ is the same, as discussed in the text.}
\label{QPtYUncertELoss}
\end{figure}

\begin{figure}
\includegraphics[width=0.48\textwidth]{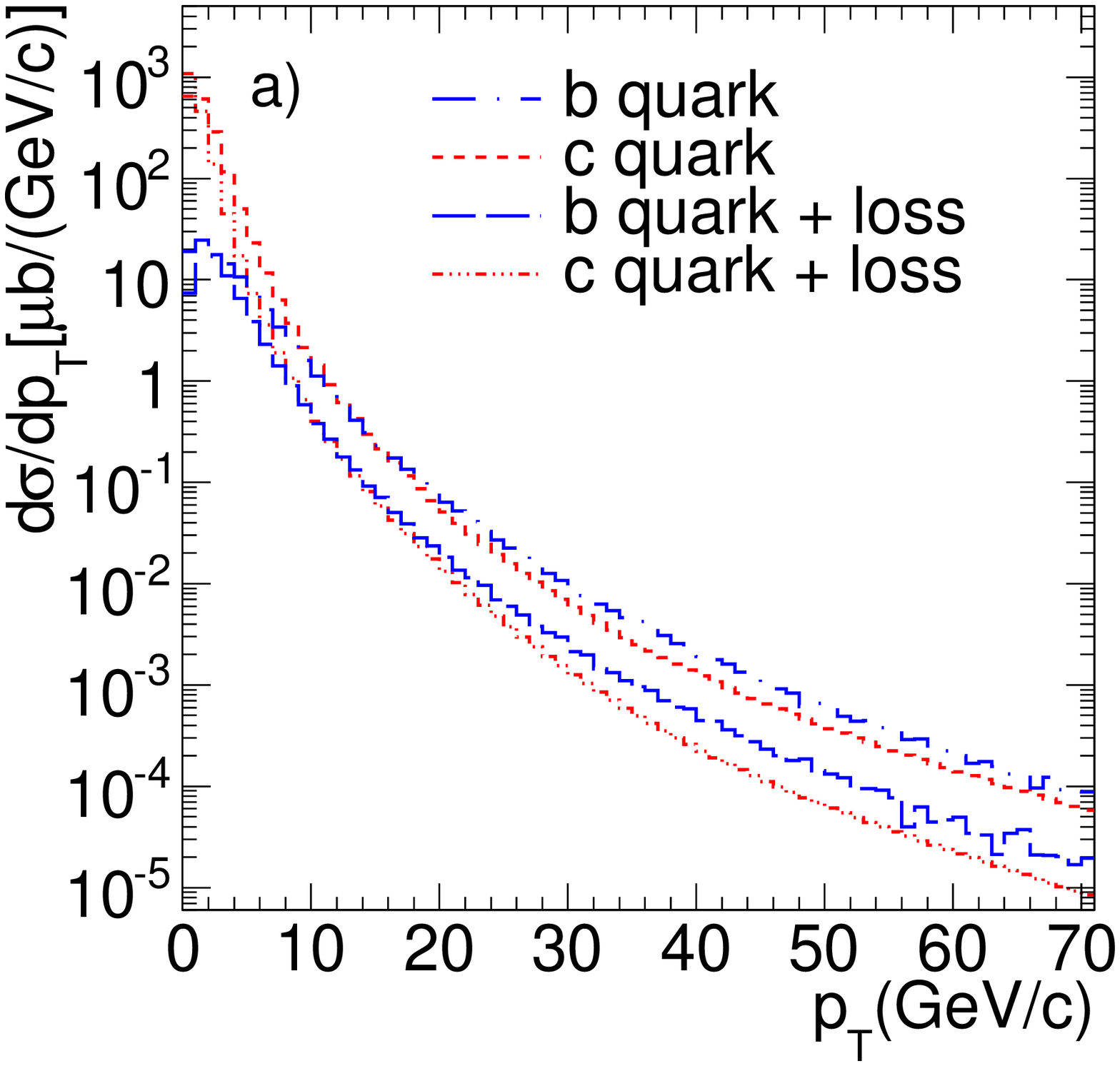}
\includegraphics[width=0.48\textwidth]{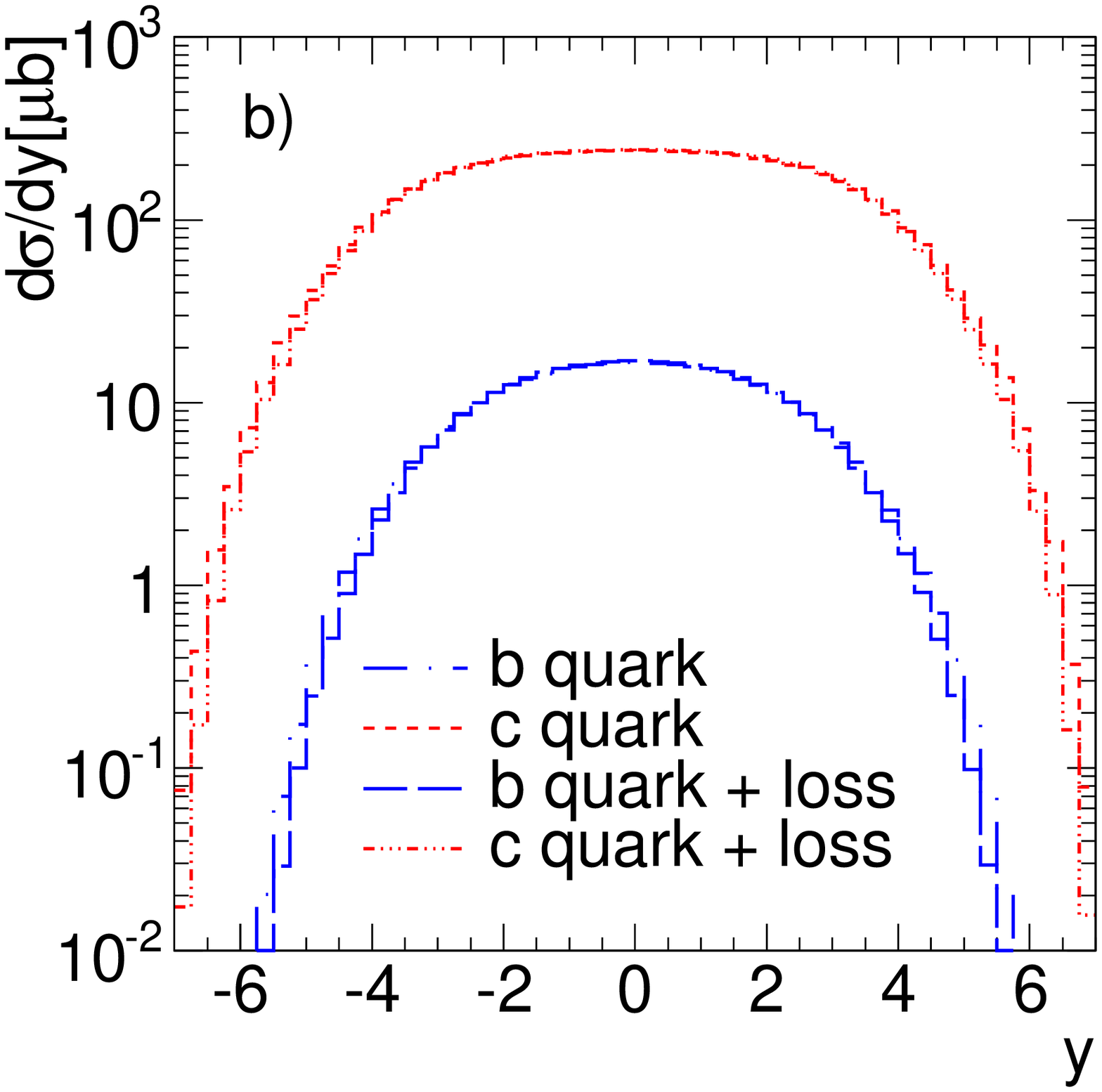}
\caption{(Color online) The inclusive single charm and bottom quark 
per nucleon cross sections as a function of $p_T$ (left) and rapidity (right) 
both with and without energy loss in Pb+Pb collisions at 
$\sqrt{s_{_{NN}}}= 2.76$ TeV. The cross sections, given per nucleon, include 
modification of the initial
parton distributions via the central EPS09 shadowing parameterization.}
\label{charm}
\end{figure}

We assume that all the observed heavy flavor production in Pb+Pb collisions
occurs during the initial nucleon-nucleon collisions.  Thermal production of
$Q \overline Q$ pairs is expected to be only a fraction of this initial
production \cite{GAVIN} unless the plasma is composed of massive quasi-particles
which would lower the effective threshold for heavy flavor production in the
medium \cite{LEVAI}, enhancing production in this channel.  However, such
production would be at lower transverse momentum and with a narrower rapidity
distribution than shown in Fig.~\ref{charm}.

\begin{table}
\caption[]{Heavy flavor and Drell-Yan cross sections at 
$\sqrt{s_{_{NN}}}= 2.76$ TeV.  The cross sections are given per nucleon while
$N_{Q \overline Q}$ and $N_{l^+ l^-}$ are the number of $Q \overline Q$ and lepton 
pairs per Pb+Pb event.  The uncertainties in the heavy flavor cross section are
based on the Pb+Pb central values with the mass and scale uncertainties added
in quadrature.}
\label{NLOcros}
\begin{tabular}{|c|c|c|c|} 
\hline 
                 & $ c \overline c$  &  $b \overline b$     & DY\\
                 &              &          & $1 \leq M \leq 100$ GeV\\
\hline
$\sigma_{\rm PbPb}$ & $1.76^{+2.32}_{-1.29}$ mb & $89.3^{+42.7}_{-27.2}$ $\mu$b & 70.97 nb \\
$N_{Q\overline Q}$ & $9.95^{+13.10}_{-7.30}$ & $0.50^{+0.25}_{-0.15}$ &  -      \\
$N_{\mu^+\mu^-}$  & $0.106^{+0.238}_{-0.078}$ & $0.0059^{+0.0029}_{-0.0017}$ & 0.0004  \\
\hline
\end{tabular}
\end{table} 

The heavy quarks are decayed semileptonically and lepton pairs are formed
from correlated $Q \overline Q$ pair decays.  We do not consider uncorrelated
$Q \overline Q$ contributions to the continuum since these should be
eliminated by a like-sign subtraction.  We assume that any uncorrelated
dileptons from $c \overline b$ and $\overline c b$ decays are also removed
by like-sign subtraction and that lepton pairs from a single chain decay,
$B\rightarrow D l_1 X \rightarrow l_1 l_2 X'$, only contribute to the low mass
continuum, see Ref.~\cite{LIN2}.  The number of lepton pairs is obtained from
the number of $Q\overline Q$ pairs,
\begin{eqnarray}
N_{\mu^+\mu^-} = N_{Q \overline Q} [B(Q \rightarrow l X)]^2 \, \, .
\end{eqnarray}
The values of $N_{Q \overline Q}$ and $N_{\mu^+ \mu^-}$ are given in 
Table~\ref{NLOcros}, along with their uncertainties.

Dilepton production by the Drell-Yan process has also been calculated
to NLO in pQCD \cite{vanNeerven}.  The cross section in the mass interval
$1 < M < 100$ GeV, including EPS09 shadowing in Pb+Pb collisions, is given
in Table~\ref{NLOcros}.  The integrated cross section is dominated by the lowest
masses.  The largest potential modification due to the presence
of the nucleus is on the low mass rate, in the resonance region.  At larger 
masses, this effect becomes competitive with the effects of the relative number
of protons and neutrons in the nucleus compared to a $pp$ collision (isospin
effects) \cite{HPpA}.  We have used PYTHIA \cite{PYTHIA} to generate the 
Drell-Yan $p_T$ distribution and to place kinematic cuts on the individual 
leptons of the pair.  The total rate has been normalized to the calculated
NLO cross 
section.  The pQCD uncertainties on the Drell-Yan rate, particularly above the
resonance region, are not large.  In general, they are smaller than the
uncertainties due to the shadowing parameterization \cite{HPpA}.

Finally, we include energy loss effects on the charm and bottom quarks. 
Since heavy quarks do not decay until after they have traversed the
medium, their contribution to the final dilepton spectra will reflect its 
influence.  Indications from inclusive non-photonic lepton spectra at RHIC
\cite{RAARHIC}, attributed to heavy flavor decays, suggest that the effects of
energy loss are strong and persist up to high $p_T$.  They also suggest that
the magnitude of the loss is similar for that of light flavors, {\it i.e.}
independent of the quark mass so that the effects are similar for charm and
bottom.  The source of this loss as well as its magnitude are still under 
investigation, see Ref.~\cite{FUV} and references therein.

To estimate the effects of energy loss on the dilepton continuum, we adjust the
heavy quark fragmentation functions to give a value of $R_{AA}$ for each flavor
separately that is consistent with the measured prompt lepton $R_{AA}$
in central Pb+Pb collisions at high $p_T$, $R_{AA} \sim
0.25 - 0.30$ \cite{RAAAlice}, for both charm and bottom quarks.  We then use
these modified fragmentation functions to calculate the 
medium-modified dilepton distributions from heavy flavor decays.

Including energy loss does not change the total cross section since it moves
the quarks to lower momentum without removing them from the system.  Thus the
$p_T$-integrated rapidity distributions are also unaffected, see 
Fig.~\ref{QPtYUncertELoss}, which presents the single inclusive heavy flavor 
production uncertainty bands after energy loss.  
The charm and bottom quark $p_T$ distributions still
exhibit the same general behavior: the slopes are parallel to those without 
energy loss at high $p_T$ but show a pile up of low $p_T$ quarks after loss
is included.  After taking energy loss into account, the point where the bottom 
quark distribution begins to dominate is shifted to lower $p_T$, 
$\sim 10$~GeV$/c$ instead of $\sim 20$~GeV$/c$ 
when the widths of the bands are accounted for.
  
The relative strength of charm and bottom energy loss in medium is not
yet settled.  Although bottom quarks are expected to lose less energy than
charm quarks, the data from RHIC and LHC exhibit important differences
\cite{PHENIXqm12,CMSqm12}.  If we assume that bottom quarks lose less energy
than charm, then the bottom and charm quark uncertainty bands in 
Fig.~\ref{QPtYUncertELoss} will separate at high $p_T$ with the bottom quark
band above that of the charm.  

Figure~\ref{charm} compares the central values of the uncertainty bands 
with and without energy loss directly.  We note that the difference in the
heavy flavor $p_T$ distributions due to energy loss is larger than the 
uncertainty bands with and without energy loss.
The rapidity distributions do not show any significant effect due to
energy loss since the results are shown integrated over all $p_T$.  Since the
total cross sections are unchanged without any acceptance cuts, there  is an
effect only at far forward rapidity.

\section{Thermal dilepton production}

The contribution of thermal dileptons is calculated assuming that
a QGP is formed in local thermal equilibrium at some initial temperature $T_i$ 
and initial time $\tau_i$ which cools hydrodynamically through a 1D Bjorken 
expansion \cite{Bjorken}.  Assuming a first-order phase transition, when the 
QGP cools to the critical temperature 
$T_c$ at time $\tau_c$, the temperature of the system is held fixed until 
hadronization is completed at time $\tau_h$. Afterwards, the hadron gas
cools to the freeze-out temperature $T_f$ at time $\tau_f$ \cite{KAJA}.

The thermal dilepton emission rate due to $q \overline q \rightarrow l^- l^+$
is \cite{KAJA,VOGT} 
\begin{eqnarray}\label{quarkrate}
 {dN \over d^4x d^2p_T dy dM^2} 
&=& \frac{3}{(2\pi)^5} M^2 \sigma(M^2) F
        \exp(-E/T) \nonumber \\
&=& \frac{\alpha^2}{8\pi^4} F 
        \exp(-E/T) \, \, .
\end{eqnarray}
Here $M$, $p_T$ and $y$ are the mass, transverse momentum,
and  rapidity of the lepton pair while $d^4x = \tau d\tau \eta \pi R_A^2$ where
$\eta$ is the rapidity
of the fluid with temperature $T$ and $R_A = r_0 A^{1/3}$.
The mass-dependent cross section, $\sigma(M^2) = F~4\pi\alpha^2/3M^2$ includes a
factor $F$ that depends on the phase of the matter.
In a two-flavor QGP, $F_{\rm QGP} = \sum e_q^2 = 5/9$, while, in the hadronic 
phase, form factors representing the resonance region \cite{GALE} are used.  
We concentrate on masses above the resonance region.
In the mixed phase,
\begin{eqnarray}
F = (1-h(\tau)) \, F_{\rm QGP}
   + h(\tau) \,  F_{\rm had} \, \, ,
\end{eqnarray}
where $h(\tau)$ is the hadron fraction of the mixed phase.

The dilepton $p_T$ distribution is 
\begin{eqnarray}\label{EqnPt}
\frac{dN}{d^4x dy dM dp_T } = \frac{\alpha^2}{4\pi^4}  F \, M \, p_T \,
   \exp\left(-\frac{ \sqrt{M^2 + p_T^2}  \cosh (y-\eta)}{T}\right) \, 
\end{eqnarray}
and the dilepton invariant mass distribution, integrated over $p_T$, is
\begin{eqnarray}\label{EqnM}
\frac{dN}{d^4x dy dM} = \frac{\alpha^2}{2\pi^3} F \, M^3 \,
\left({1\over x^2} + {1\over x}\right) \exp(-x),
\end{eqnarray}
where
\begin{eqnarray}
x=\frac{ M \cosh (y-\eta)}{T}.
\end{eqnarray}

\begin{figure}
\includegraphics[width=0.48\textwidth]{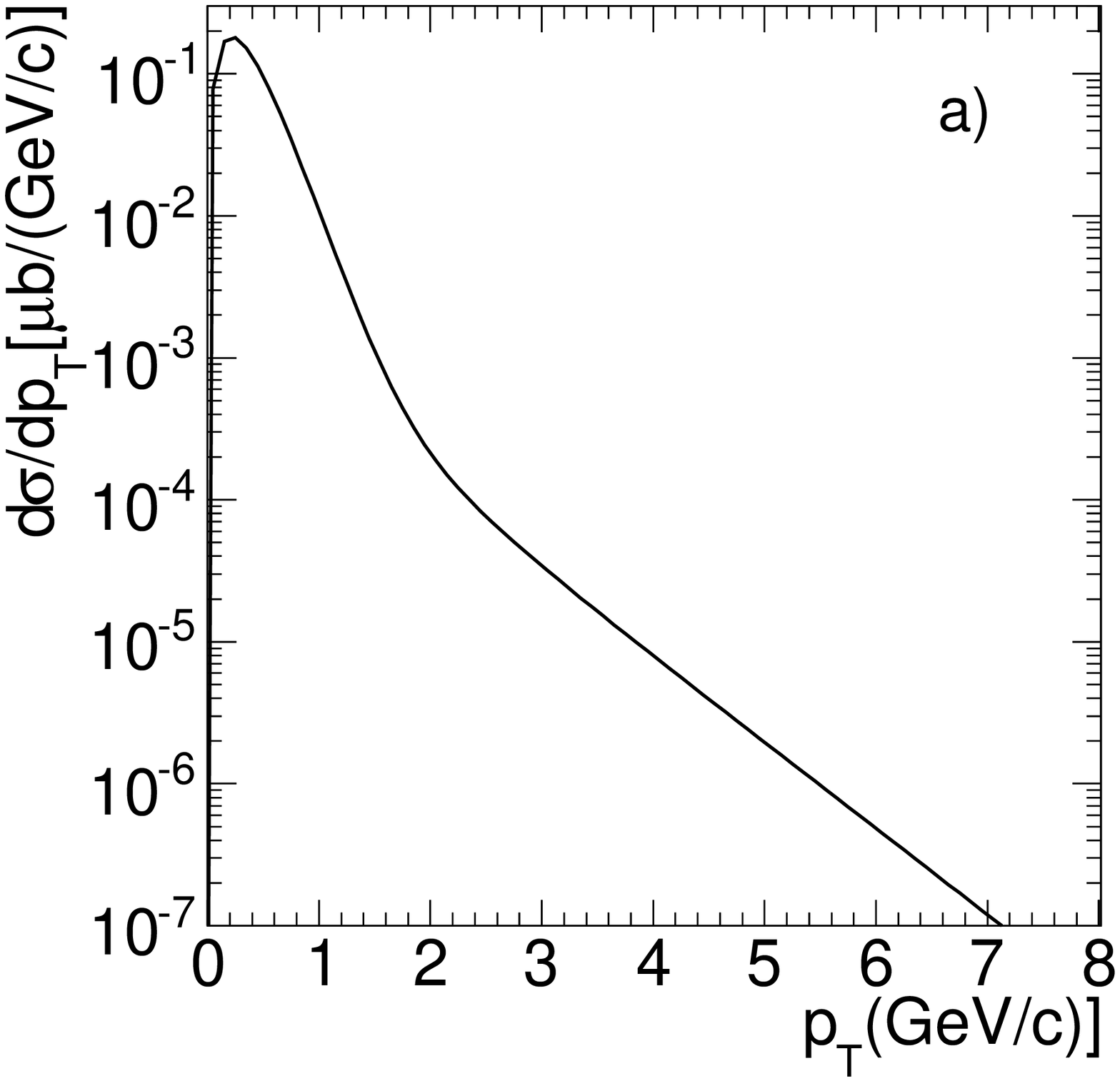}
\includegraphics[width=0.48\textwidth]{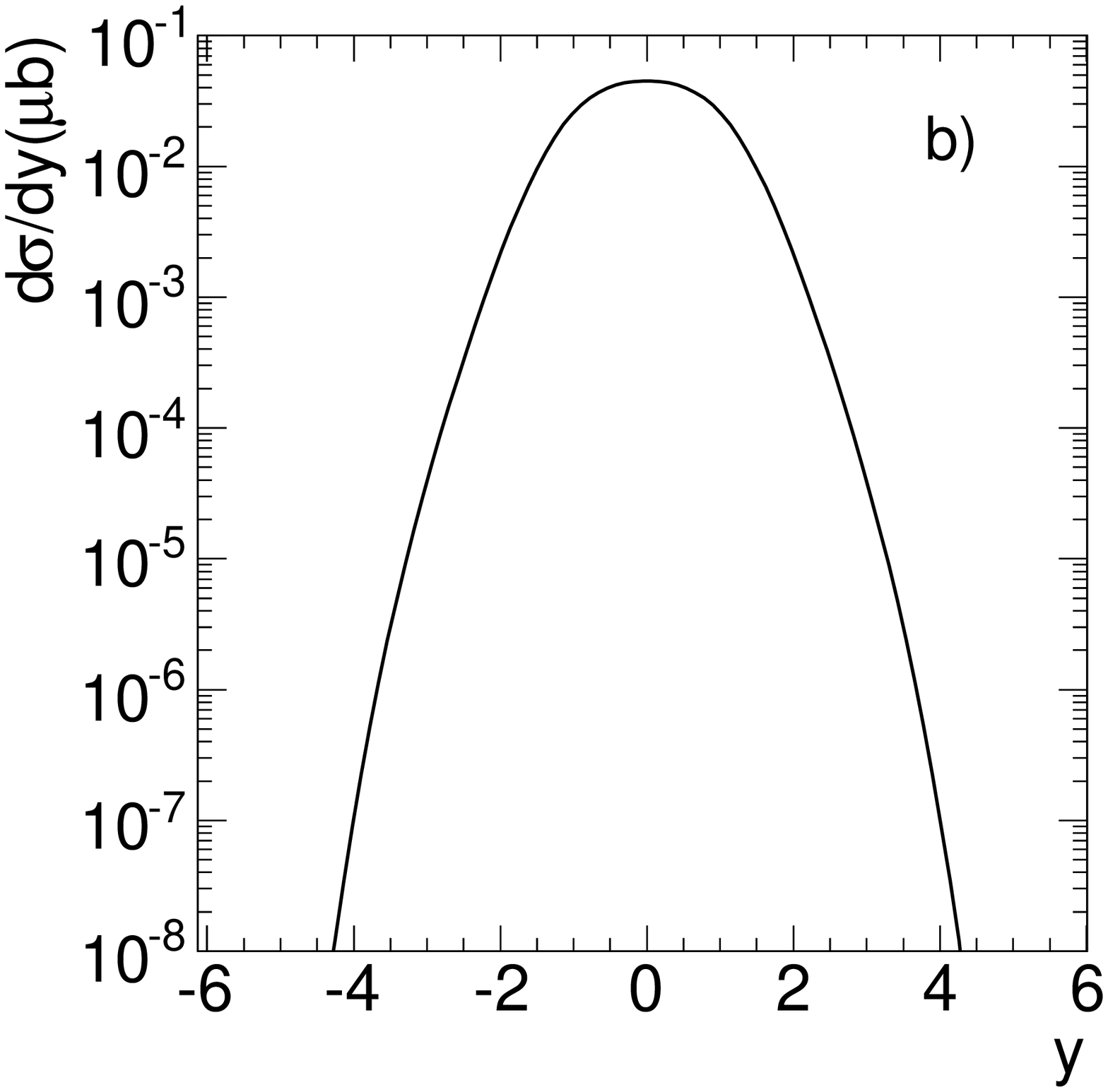}

\caption{The thermal dilepton cross section as a function of $p_T$ 
(left) and rapidity (right) in Pb+Pb collisions at $\sqrt{s_{_{NN}}}= 2.76$ 
TeV.}
\label{Thermal}
\end{figure}

The initial time is assumed to be $\tau_i = 0.1$ fm/$c$. 
The initial temperature $T_i$ is obtained from
the total multiplicity distribution,
\begin{eqnarray}
{dN \over dy} = \tau_i T_i^3 4a_q \pi R_{A}^2/3.6 \, \, ,
\end{eqnarray}
where $dN/dy=1.5~dN_{\rm ch}/dy$. The charged particle multiplicity, 
$dN_{\rm ch}/dy = 1600$, was measured in Pb+Pb collisions 
at 2.76 TeV \cite{MULT}.  Using this value with $a_q = 37 \pi ^2 /90$ 
gives $T_i = 636$ MeV. 
The temperature decreases in the QGP as
\begin{eqnarray}
T(\tau) = T_i \left({\tau_i \over \tau}\right)^{1/3}
\end{eqnarray}
for $\tau_i~<~\tau~<~\tau_c$.  The temperature in mixed phase is 
$T$ = $T_c$ = 160 MeV. The mixed phase ends at $\tau_h = (a_q/a_h) \tau_c$
where $a_h = 3 \pi ^2 / 90$ for a pion gas.
The hadronic fraction of the mixed phase, $h(\tau)$, is
\begin{eqnarray}
h(\tau) ={a_q \over a_q - a_h}\left( {\tau -\tau_c \over \tau} \right) \,\, .
\end{eqnarray} 
The temperature in hadron phase between $\tau_h~<~\tau~<~\tau_f$,
is 
\begin{eqnarray}
T(\tau) = T_c \left({\tau_h \over \tau}\right)^{1/3} \, \, .
\end{eqnarray}

The thermal dilepton rate given in Eqs.~(\ref{EqnPt}) and
(\ref{EqnM}) is converted to a cross section by dividing the rate by the
minimum bias nuclear overlap, $T_{\rm PbPb}$.
Figure~\ref{Thermal}(a) and (b), shows the differential cross sections for thermal
dilepton production as a function
of $p_T$ and rapidity.  The $p_T$ distribution, integrated over pair mass, shows
two slopes, a steep decrease when the minimum pair transverse mass, $M_T$, is 
on the order of the temperature and a long tail when $M_T \gg T$.  The rapidity
distribution is significantly narrower than those resulting from the initial
hard scatterings shown in Fig.~\ref{charm}. 

This simple application of a one-dimensional Bjorken expansion through a 
first-order phase transition significantly overestimates the lifetime of
the hot system. Thus, the results shown in Fig.~\ref{Thermal} should be 
regarded as an upper limit on the thermal
contribution. 

To obtain the pair mass distributions including single lepton cuts, single
leptons are generated by a Monte Carlo based on the pair $M$, $p_T$ and $y$ 
distributions using energy-momentum conservation.

\section{Results and discussion}

\begin{figure}
\includegraphics[width=0.48\textwidth]{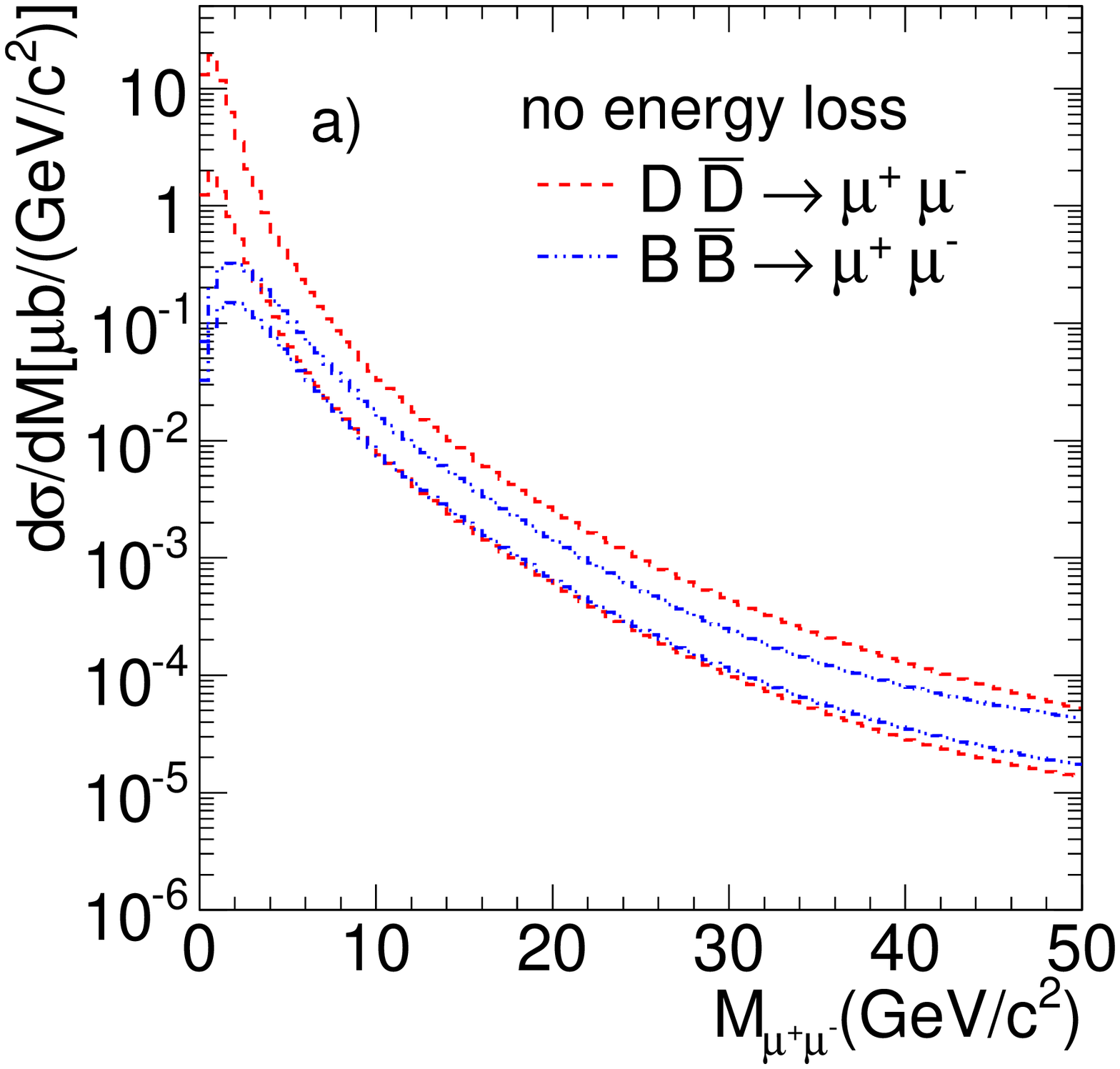}
\includegraphics[width=0.48\textwidth]{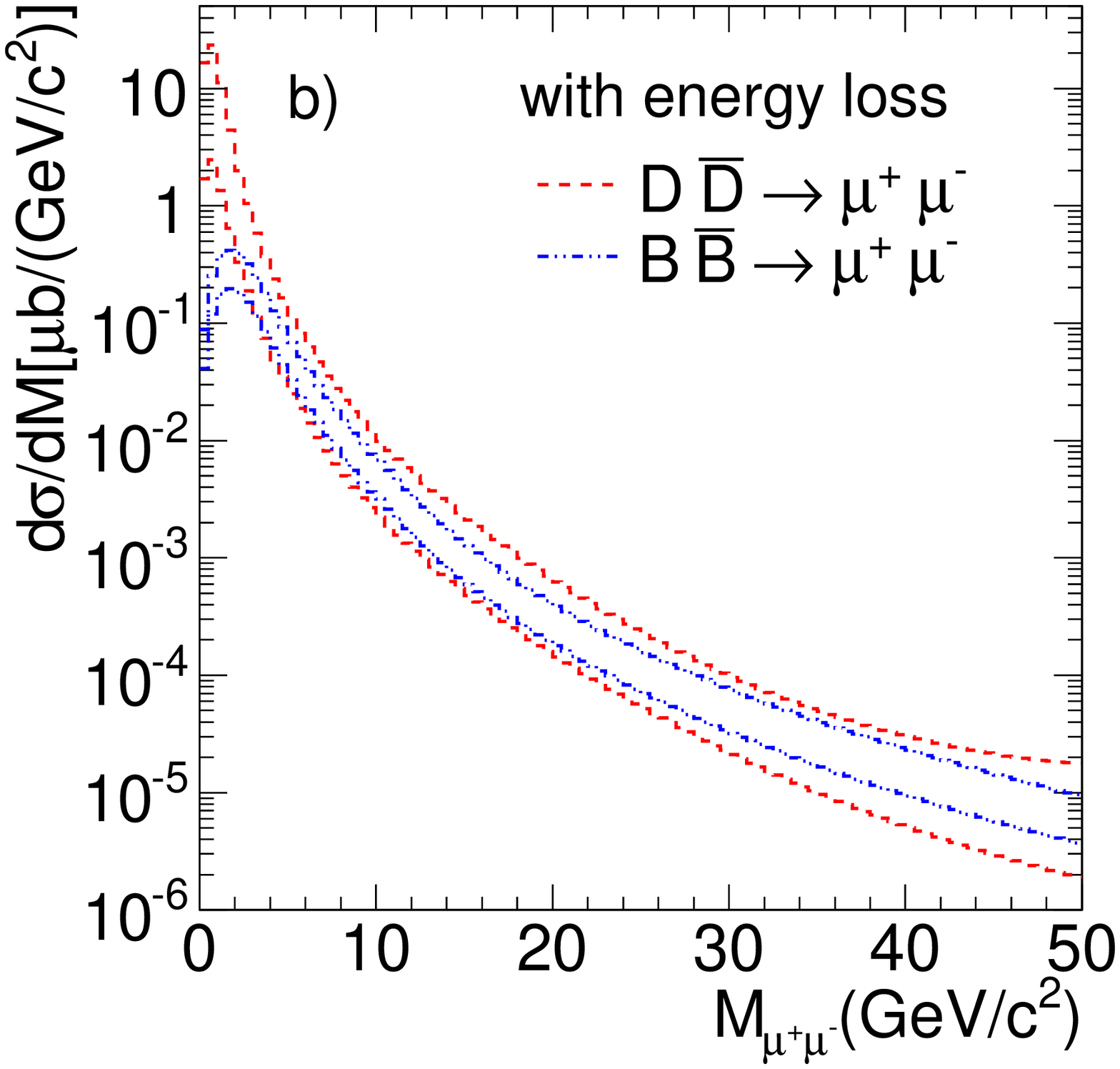}
\caption{(Color online) Theoretical uncertainty bands for the dilepton 
invariant mass distributions from semileptonic charm (red, short-dashed) 
and bottom (blue, dot-dot-dashed) decays. The uncertanities are calculated 
the same way as in Sect.~\protect\ref{harddilep}.}
\label{ScaleAndMassVariation}
\end{figure}

In Fig.~\ref{ScaleAndMassVariation}, we show the theoretical uncertainty bands 
on the dilepton invariant mass distributions from semileptonic charm and bottom 
decays. The uncertainty bands for the decay dileptons are calculated identically
to those of the charm and bottom quark distributions shown in 
Sec.~\ref{harddilep}.  The dilepton uncertainty bands are broader than those
for the single inclusive heavy flavors and, here, the dilepton band from charm
decays is wider than for bottom.  This is the case both without, 
Fig.~\ref{ScaleAndMassVariation}(a), and with, 
Fig.~\ref{ScaleAndMassVariation}(b), energy loss.  
While we show only the central values of these distributions in the remainder 
of this section, it is important to keep in mind the significant mass and 
scale uncertainties in heavy flavor production, considerably larger
than those on high mass Drell-Yan production.

\begin{figure}
\includegraphics[width=0.48\textwidth]{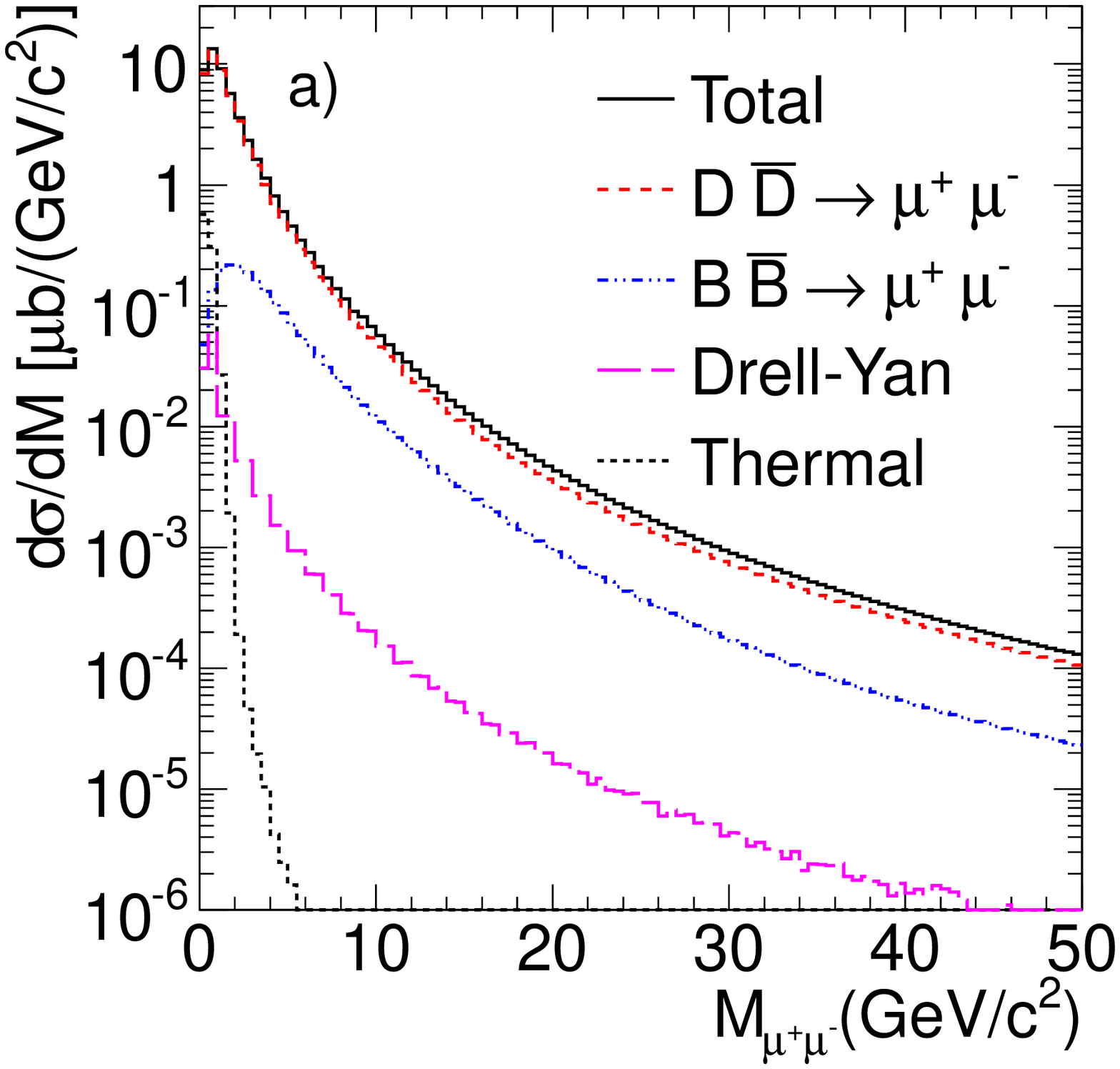}
\includegraphics[width=0.48\textwidth]{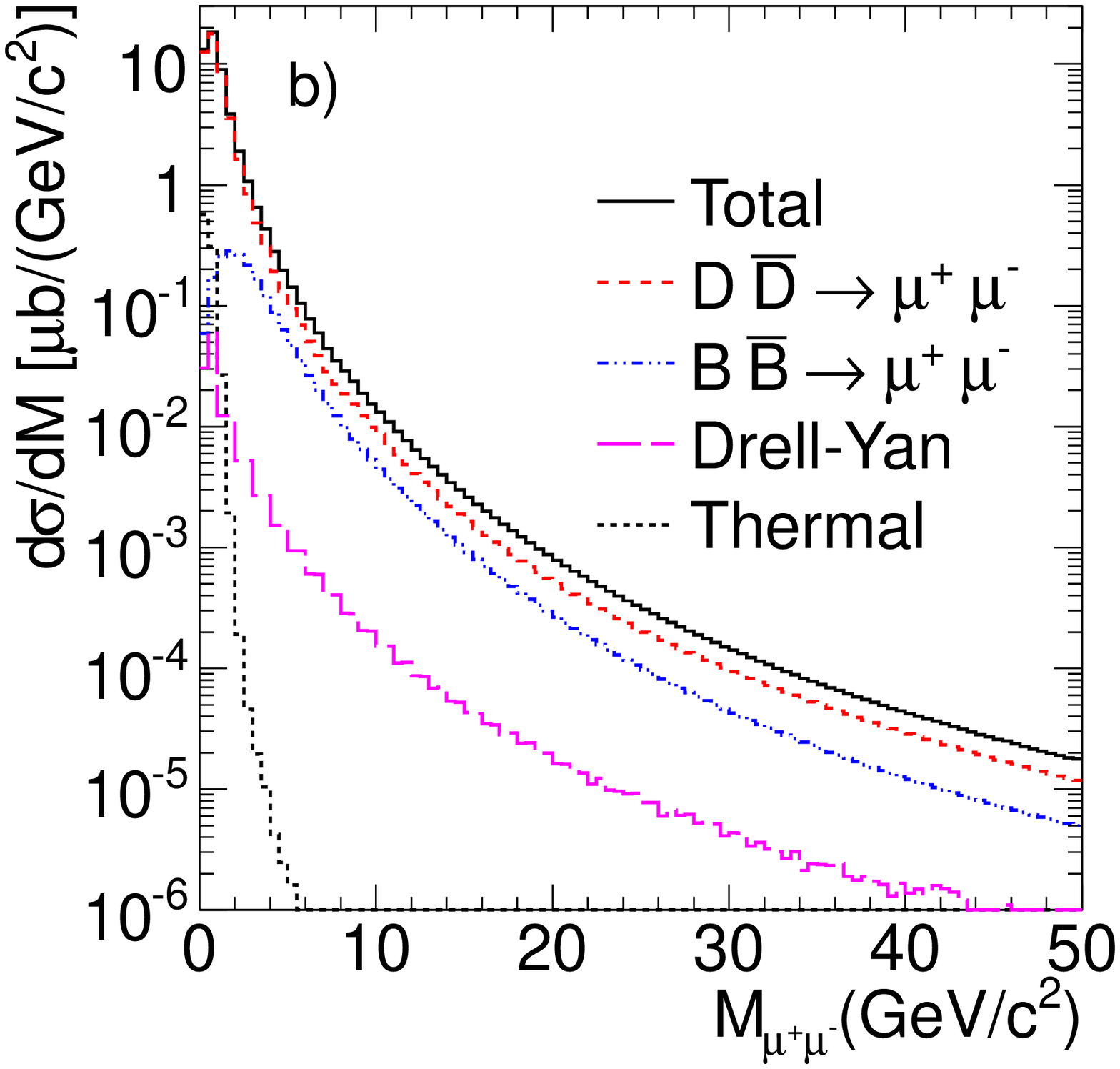}

\caption{(Color online)
The invariant mass distributions for the four contributions to the 
dilepton spectra discussed here: semileptonic charm (red, short-dashed) 
and bottom (blue, dot-dot-dashed) decays, Drell-Yan (magenta, 
long-dashed) and thermal (black, dotted) dileptons along with the sum (black,
solid) in Pb+Pb collisions per nucleon pair
at $\sqrt {s_{NN}}=2.76$ TeV. Left pannel shows distributions without any
final state energy loss, right pannel is after including heavy quark energy loss 
in the medium. The per nucleon cross sections are given.
No phase space or kinematic cuts are introduced.
}
\label{DiLepAll}
\end{figure}

Figure~\ref{DiLepAll} shows the dimuon invariant mass distributions from each
of the four sources considered: semileptonic decays of correlated $Q \overline
Q$ pairs and direct production of Drell-Yan and thermal dileptons in
Pb+Pb collisions at $\sqrt {s_{_{NN}}}=2.76$ TeV. Figure~\ref{DiLepAll}(a) shows 
the heavy flavor mass distributions without any
final-state energy loss while energy loss is included in the heavy flavor
distributions on Fig.~\ref{DiLepAll}(b).  Only the central values of the heavy
flavor contributions are shown.
The Drell-Yan and thermal dilepton
distributions are unchanged.  No kinematic cuts are included.  Without cuts, 
dileptons from $D \overline{D}$ decays dominate over the entire
mass range due to the large $c \overline{c}$ production cross section. 
Bottom pair decays are the next largest contribution followed by Drell-Yan
production.  At masses below 3 GeV/$c^2$, the Drell-Yan and thermal dilepton
contributions are competitive.  Otherwise, the thermal contribution is 
negligible. Including energy loss steepens the slope of the heavy flavor mass
distributions and also moves the $D \overline D$ decay distributions closer to
the $B \overline B$ decay distributions.  In the remainder of this section, we
will show only results with final-state heavy flavor energy loss included.

We now examine these distributions in the kinematic regimes appropriate for the
LHC detectors. CMS \cite{CMS} and ATLAS \cite{ATLAS} have excellent muon 
detectors with similar coverage  in the central rapidity region, 
$|\eta^{\mu}| \leq 2.4$. However, due to the large magnetic fields, only muons
above a rather high minimum $p_T$, $p_T > 3.0$~GeV/$c$, make it into the muon 
detectors.  ALICE \cite{ALICE} has muon acceptance on one side of the forward 
rapidity region, $2.5 \leq \eta^{\mu} \leq 4.0$.  At central rapidities,
$|\eta^{\mu}| \leq 1.0$, ALICE has an electron detector.  Some previous studies
of Pb+Pb collisions at 5.5 TeV, using leading order calculations of heavy quark
production and assuming significantly higher initial temperatures than employed 
here, suggested that thermal dileptons could be extracted from the QGP  
\cite{GALL}. Thus they reached different conclusions about the 
relative contributions of thermal and heavy flavor dileptons to the continuum.

\begin{figure}
\includegraphics[width=0.48\textwidth]{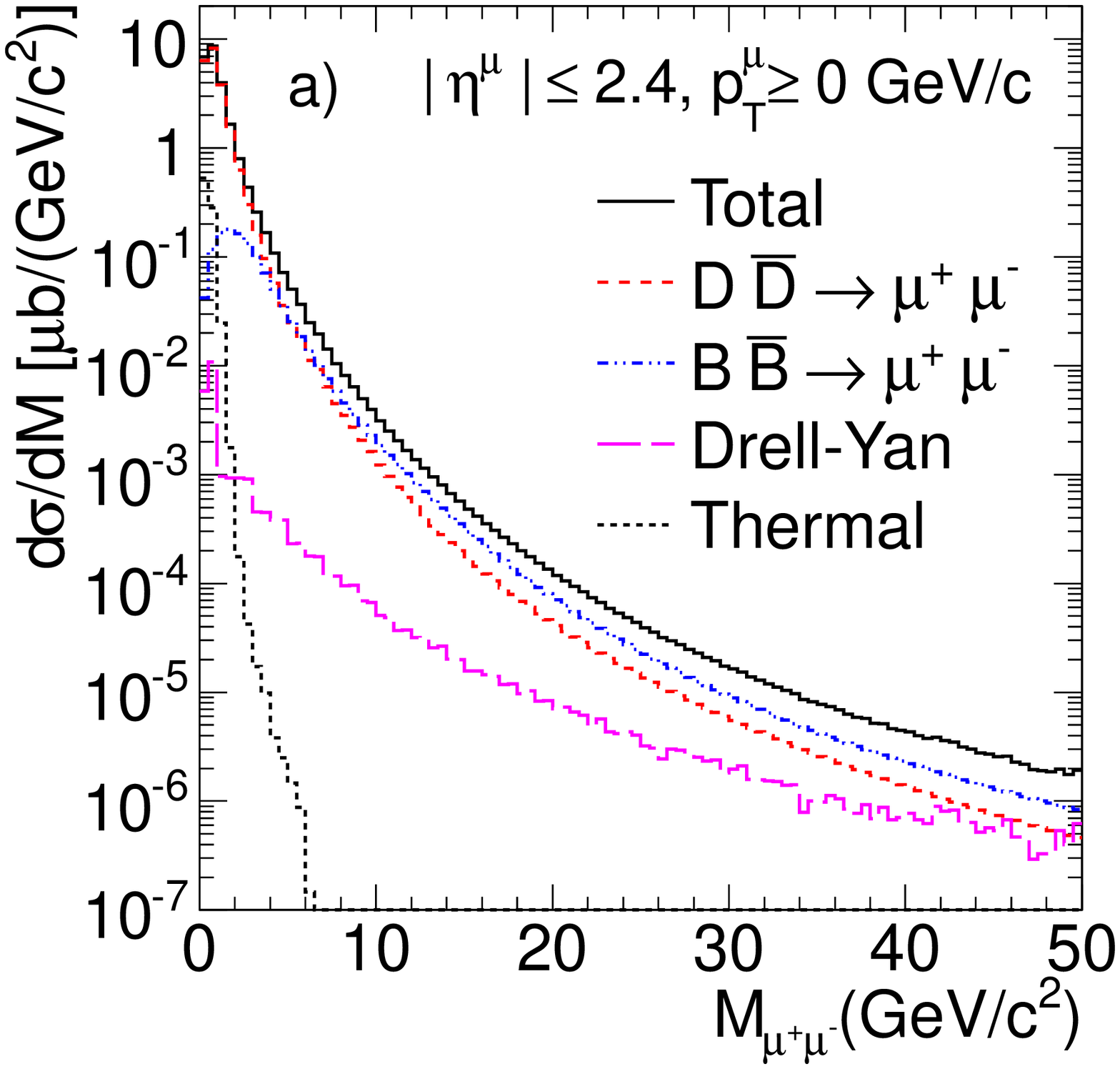}
\includegraphics[width=0.48\textwidth]{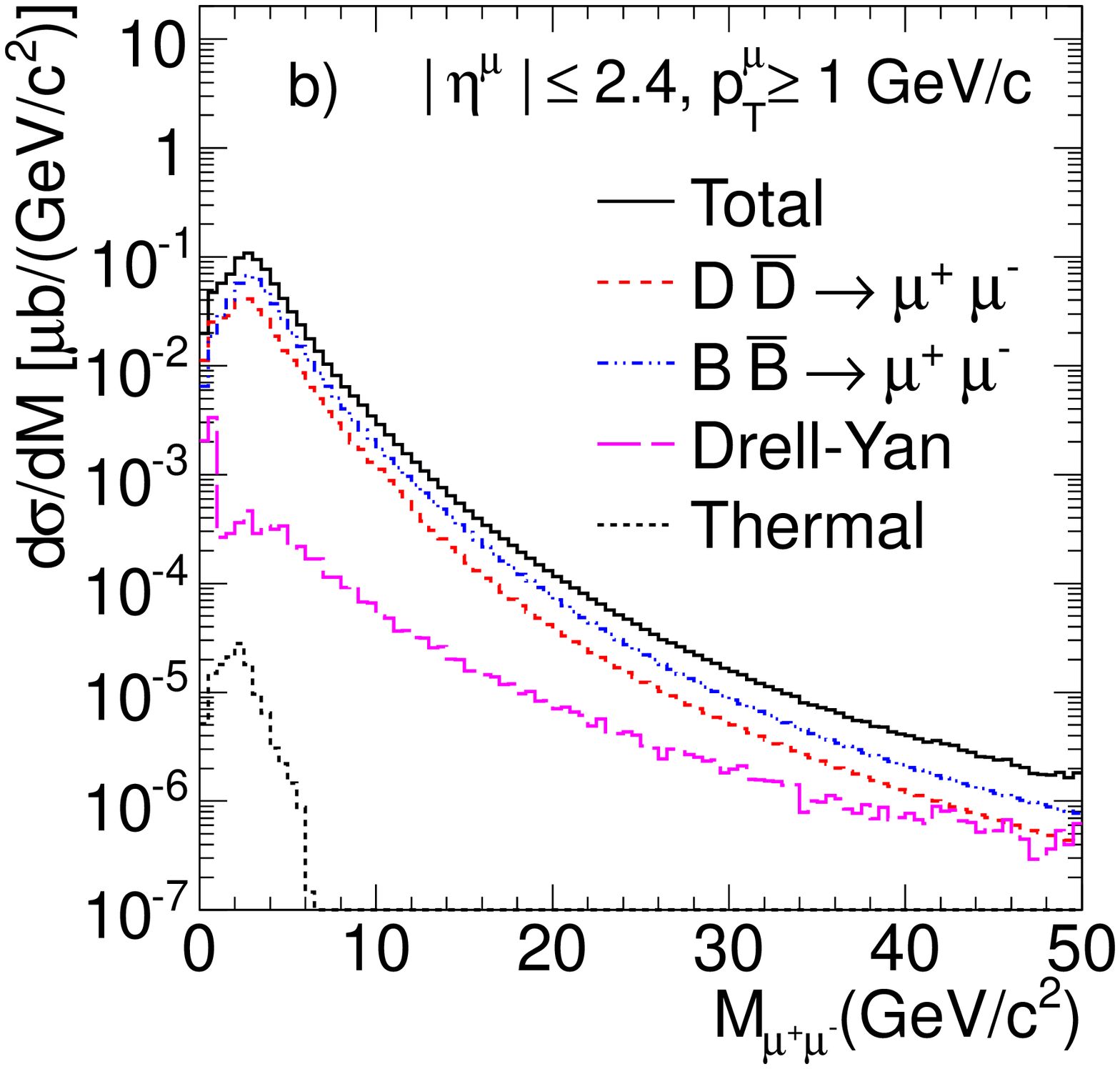}
\includegraphics[width=0.48\textwidth]{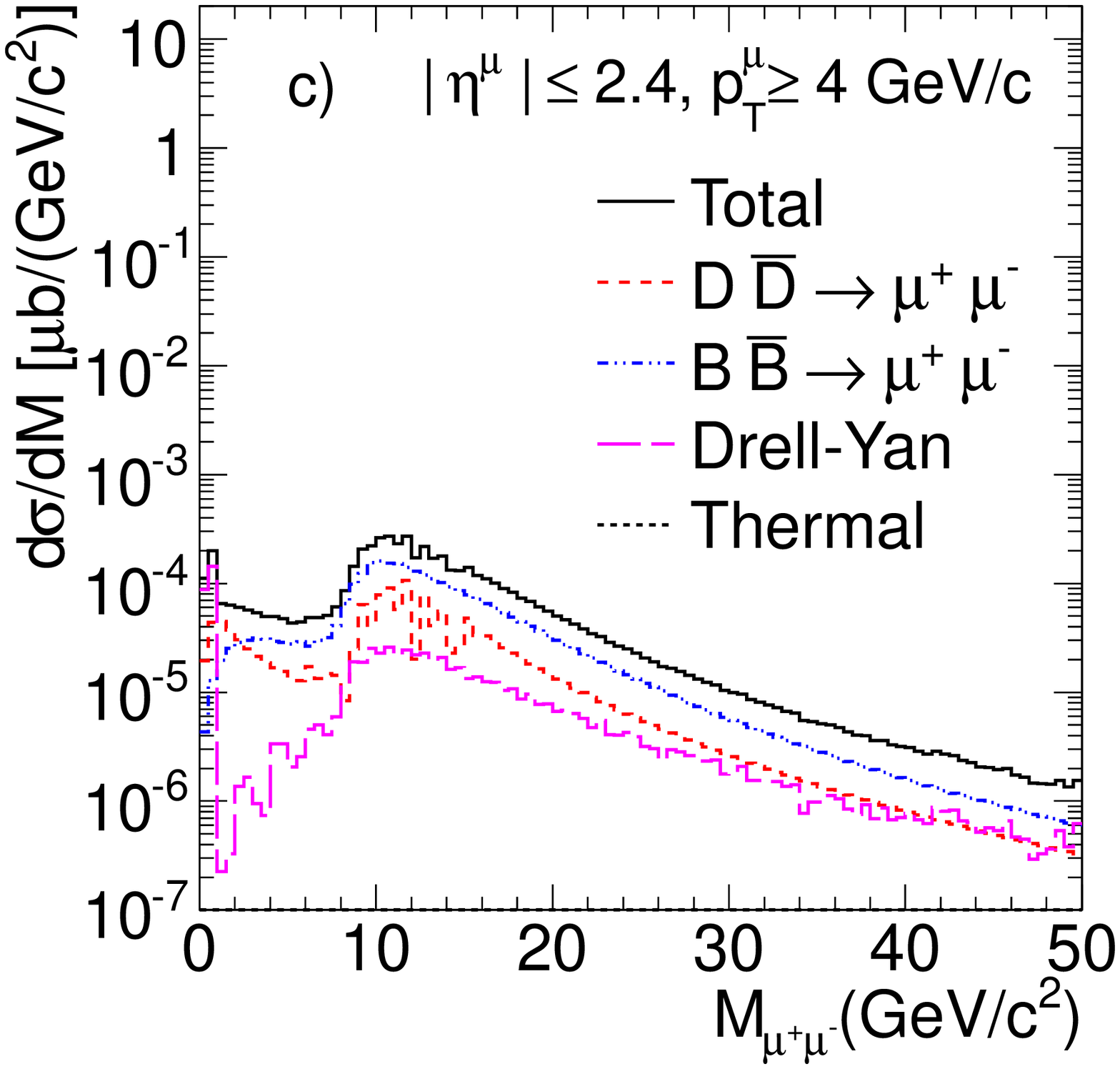}
\includegraphics[width=0.48\textwidth]{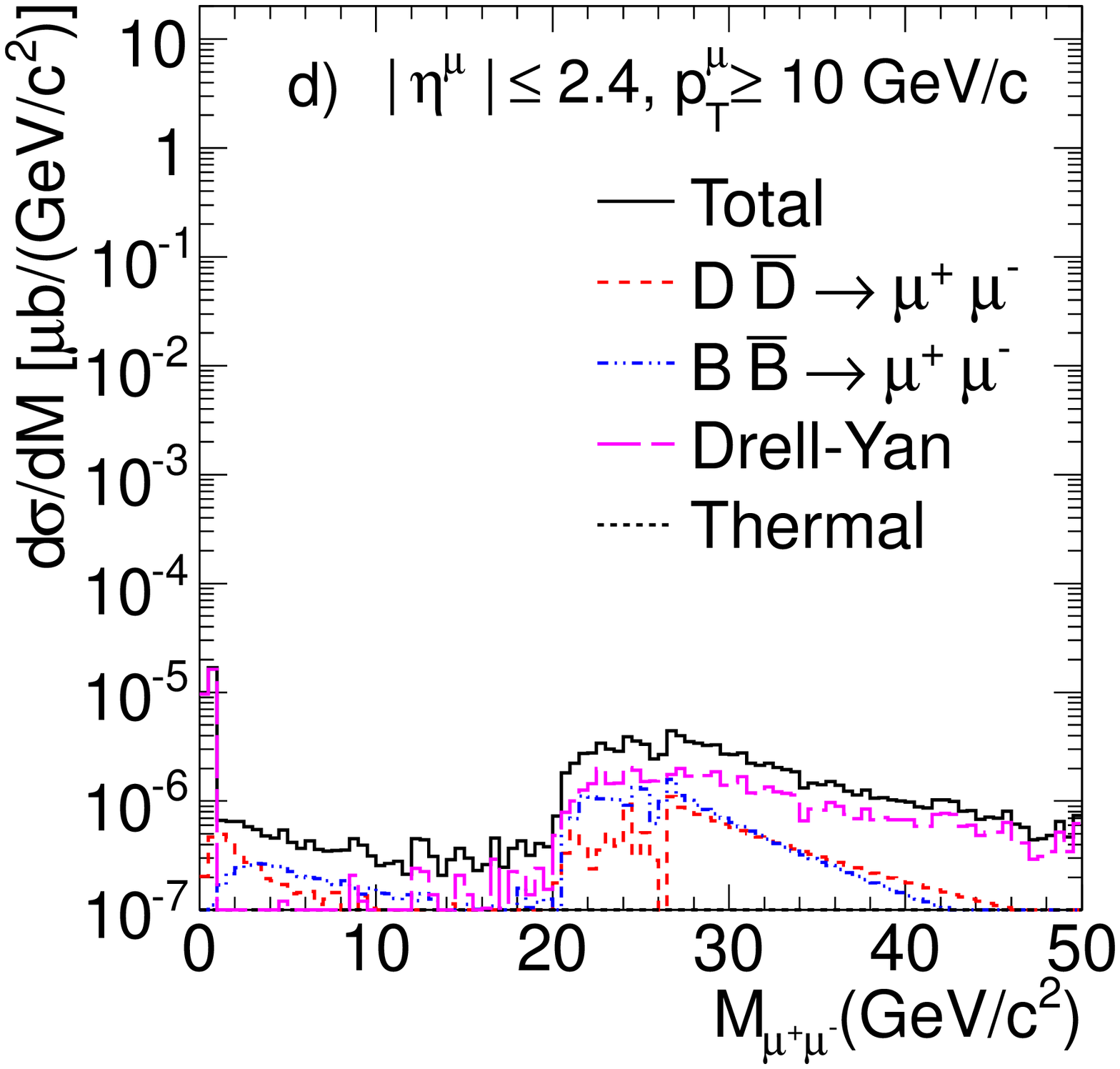}
\caption{(Color online) The same as Fig.~\protect\ref{DiLepAll} 
but now with single muon
rapidity cuts of $|\eta^\mu| \leq 2.4$.  A minimum single lepton transverse 
momentum cut of $p_T^\mu \geq 0$ (a), 1 (b), 4 (c) and 10 (d) GeV$/c$ is also 
shown.}
\label{DiLepCMST}
\end{figure}

\begin{figure}
\includegraphics[width=0.48\textwidth]{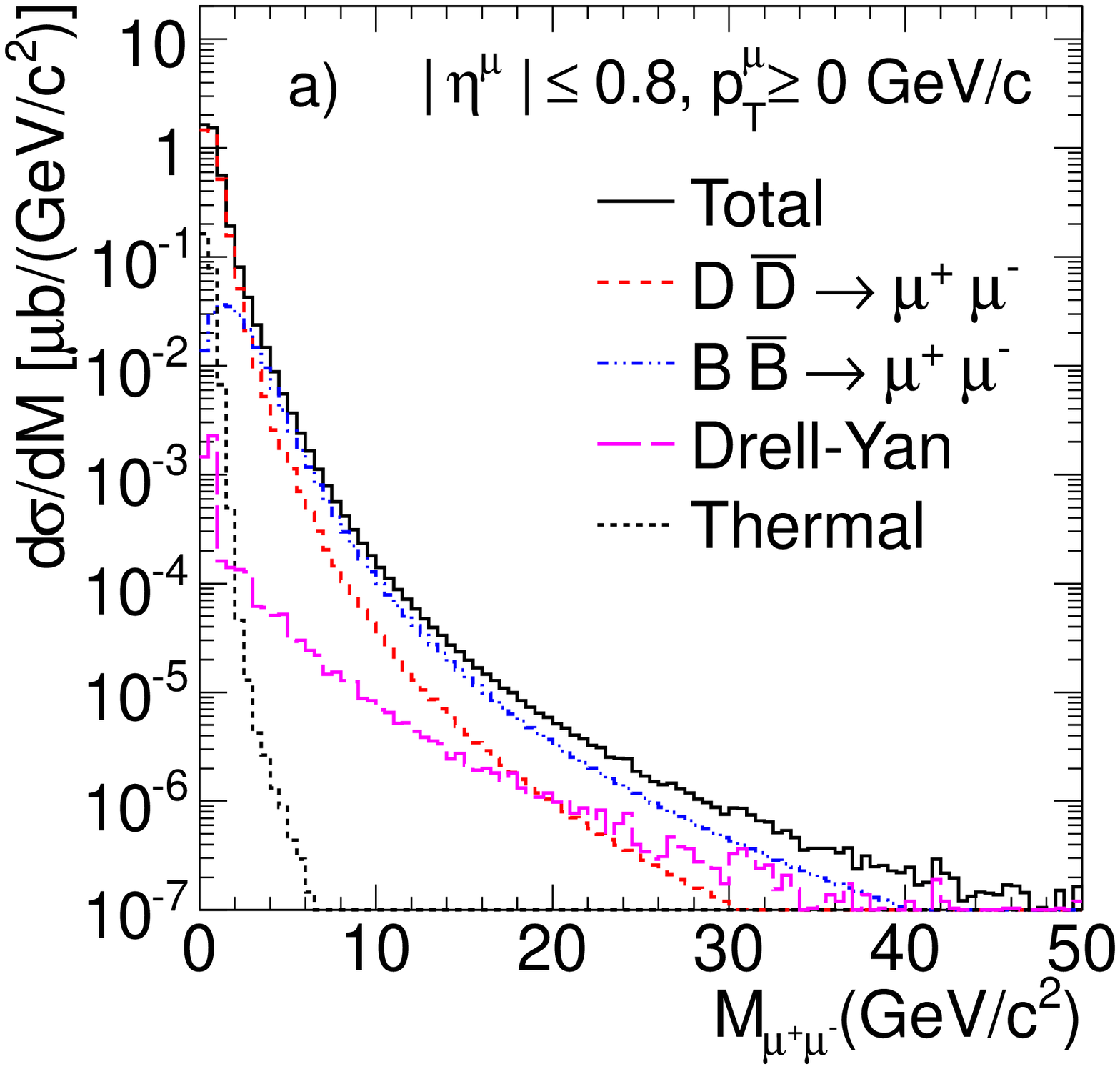}
\includegraphics[width=0.48\textwidth]{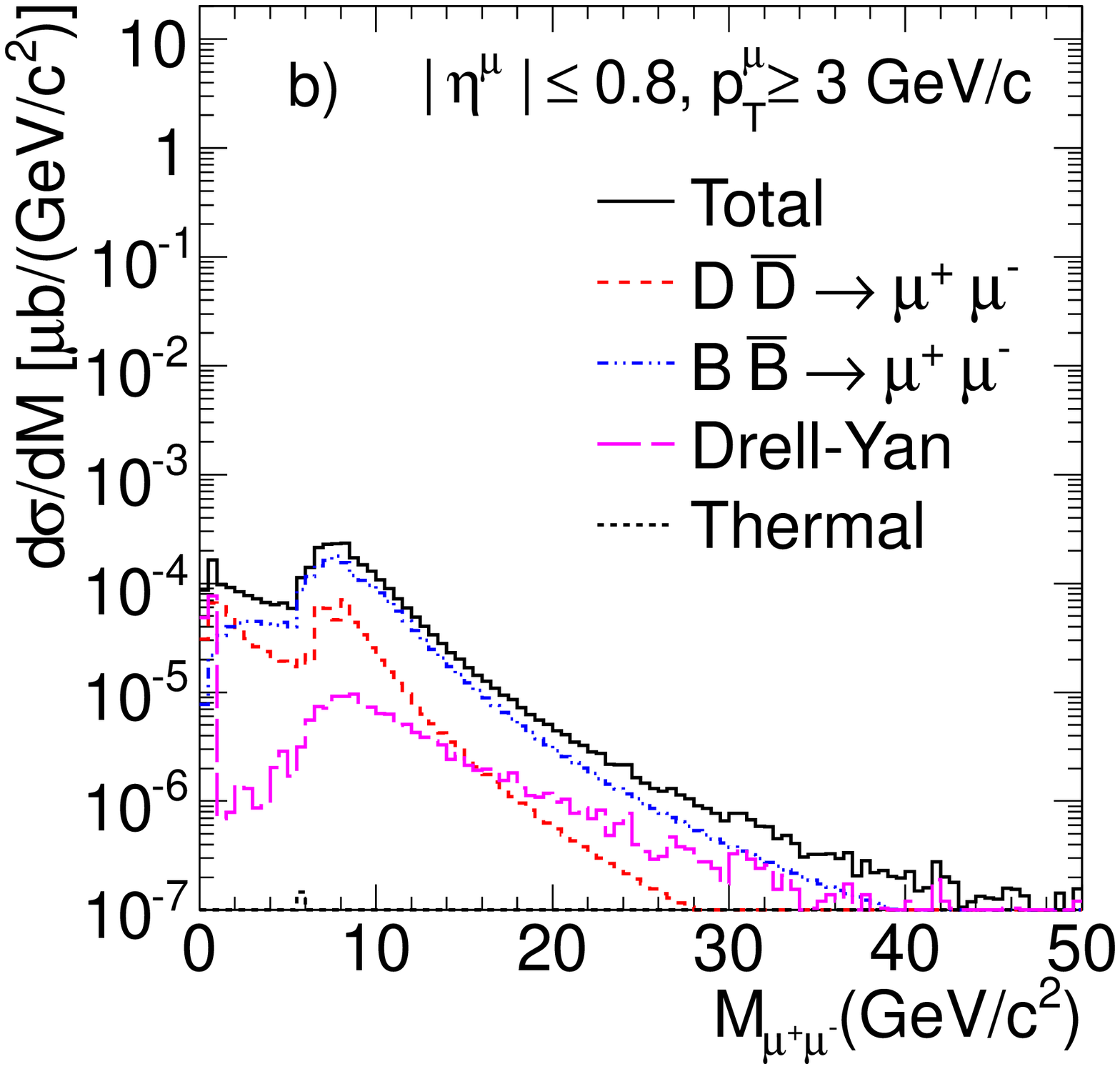}
\caption{(Color online) The same as Fig.~\protect\ref{DiLepAll} 
but now with single muon
rapidity cuts of $|\eta^\mu| \leq 0.8$.  A minimum single lepton transverse 
momentum cut of $p_T^\mu \geq 0$ (a) and 3 (b) GeV$/c$ is also shown.}
\label{DiLepCMSB}
\end{figure}

\begin{figure}
\includegraphics[width=0.48\textwidth]{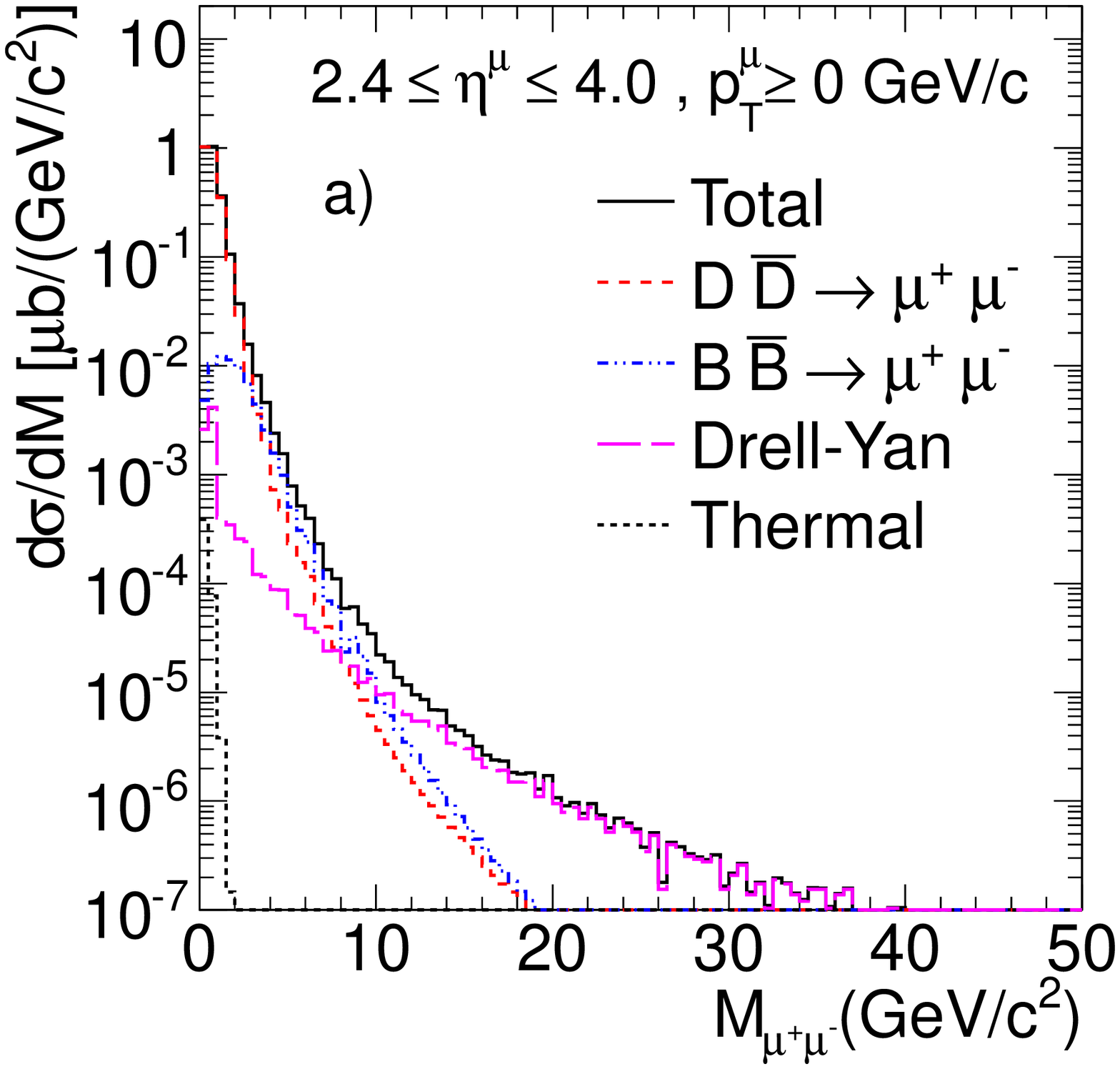}
\includegraphics[width=0.48\textwidth]{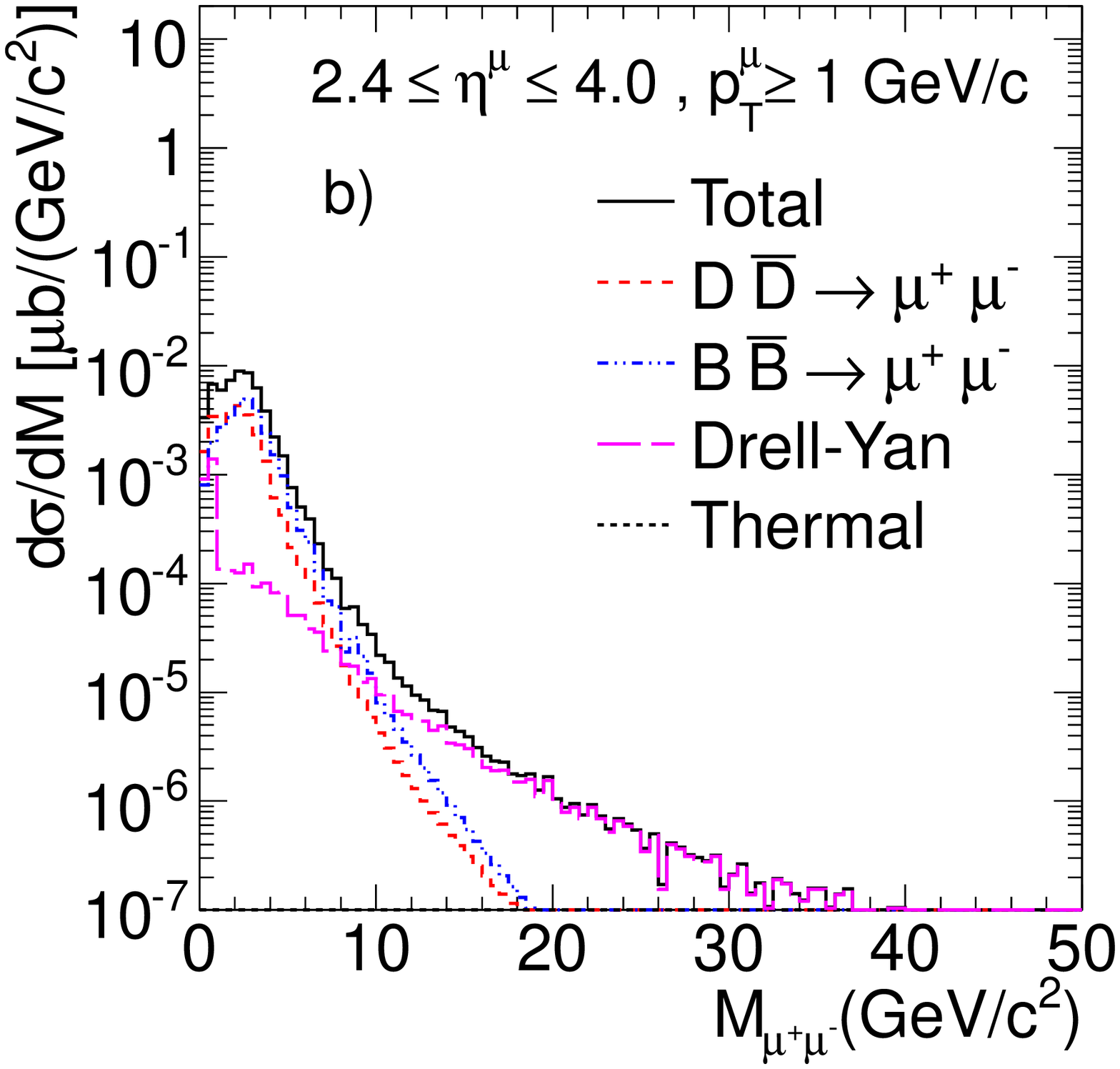}
\caption{(Color online) The same as Fig.~\protect\ref{DiLepAll} 
but now with single muon
rapidity cuts of $2.4 \leq |\eta^\mu| \leq 4$.  
A minimum single lepton transverse 
momentum cut of $p_T^\mu \geq 0$ (a) and 1 (b) GeV$/c$ is also shown.}
\label{DiLepALI}
\end{figure}

Figure~\ref{DiLepCMST} shows the dimuon invariant mass distribution for single
muons in the range $|\eta^{\mu}|\leq 2.4$, together with several muon $p_T$ 
cuts. Figure~\ref{DiLepCMST}(a) has no muon $p_T$ cut, only the $\eta$ cut.
Comparison with Fig.~\ref{DiLepAll} shows that the thermal dilepton contribution
is almost unaffected since its rapidity distribution is sufficiently narrow 
to fit within the CMS rapidity acceptance.  Since the Drell-Yan rapidity
distribution narrows with increasing mass, only the low mass region is affected
by the rather broad rapidity cut of  $|\eta^{\mu}|\leq 2.4$.  Because the charm
rapidity range is broader than that of bottom production, the dileptons from
charm decays are most affected by the rapidity cut. For $M_{\mu^+ \mu^-} > 
5$~GeV/$c^2$, the charm dilepton yield has dropped below that of bottom.

Adding a cut on single lepton $p_T$ disproportionally affects the low mass 
part of the continuum.  As the minimum lepton $p_T$ is increased from 1 GeV/$c$ 
to 10 GeV/$c$ in Figs.~\ref{DiLepCMST}(b)-\ref{DiLepCMST}(d), an ever-deepening 
dip appears in the dilepton mass distribution for $M_{\mu^+ \mu^-} < 2 p_T^\mu$. 
Even a relatively low $p_T$ cut essentially eliminates the thermal dilepton
contribution since these leptons have a rather soft $p_T$ distribution.
Since the charm and bottom quark $p_T$ distributions have the same slope for
$p_T > 7$ GeV/$c$, 
their decays are affected the same way by the lepton $p_T$ cut.
Finally, the single lepton cut of $p_T^\mu > 
10$~GeV/$c$, published with the CMS $Z^0$ measurement \cite{zboson}, based on 
approximately 50 million events, had a very low continuum 
background. This is in agreement with the result in Fig.~\ref{DiLepCMST}(d) 
which
shows that, with energy loss included, the Drell-Yan process is now the
dominant contribution to the continuum.

Figure~\ref{DiLepCMSB} shows the dimuon mass distribution in the narrower
central rapidity interval, $|\eta^{\mu}|\leq 0.8 $, equivalent to the muon 
acceptance in
the CMS barrel region and similar to the ALICE electron acceptance, 
$|\eta^{e}|\leq 1.0$.  Figure~\ref{DiLepCMSB}(a) shows the dimuon distribution 
before any $p_T$ cut. In this case, the mass distribution is more steeply 
falling in all cases except for thermal dilepton production because of its 
narrow rapidity distribution.  Since the heavy flavor hadrons decay 
isotropically to leptons, the rapidity distribution for lepton pairs is rather 
broad with a width
that is not strongly dependent on the pair mass.  Thus the narrower rapidity
acceptance reduces the high mass yields substantially relative to 
Fig.~\ref{DiLepCMST}, even before any single lepton $p_T$ cuts.  Adding a single
lepton transverse momentum cut of $p_T^\mu > 3$ GeV/$c$, Fig.~\ref{DiLepCMSB}(b),
suppresses the low mass part of the distribution.  However, the mass 
distribution is essentially unaffected by the $p_T^\mu$ cuts for $M_{\mu^+ \mu^-} > 
8$ GeV/$c^2$.

Figure~\ref{DiLepALI} shows the dimuon mass distributions in the 
forward region, $2.5 \leq \eta^{\mu} \leq 4.0$, relevant for the ALICE muon 
arm.  In this case, after energy loss, the Drell-Yan cross section rises 
above the heavy flavor decay rate for $M_{\mu^+ \mu^-} > 10$~GeV$/c^2$.  
 The heavy flavor production kinematics favors central production, with a 
rather steep decrease in the rapidity distribution as the kinematic limit is 
approached.  There is no such constraint on the resulting lepton pairs. 
 Because the decay of the individual heavy quark is isotropic in its rest 
frame, the lepton rapidity distribution has a larger plateau region, extending 
to more forward rapidity, than the parent quark. 
However, restricting the cut to one side of midrapidity eliminates many large gap pairs 
that might survive with a broad central rapidity acceptance such as in Fig.~\ref{DiLepCMST}.
Very little remains of the thermal dilepton contribution in the forward region
due to its narrow rapidity distribution.

\section{CONCLUSIONS}

In summary, we calculate open charm and bottom production and determine 
their contributions to the dilepton continuum in Pb+Pb collisions at 
$\sqrt{s_{_{NN}}} = 2.76$ TeV with and without including heavy quark energy loss.
These rates are then compared with Drell-Yan and thermal dilepton production.
The contributions of all these sources are obtained in kinematic regions 
relevant for the LHC detectors. 
  
Since most detectors accept only high $p_T$ single leptons, thermal dileptons 
would be difficult to measure.  Heavy flavours are the dominant source of 
dileptons in most kinematic regimes, even after energy loss. 
At forward rapidity, the Drell-Yan contribution begins to dominate for
$M > 10$ GeV/$c^2$.  The effects of energy loss on the decay dileptons alters 
their acceptance, particularly for high lepton $p_T$ cuts.  In most of the 
kinematic regions considered, the $b \overline b$ decay contributions become 
larger than those of $c \overline c$ for lepton pair masses greater than 
7 GeV$/c^2$. 
 
  From the approximately 50 M events collected by CMS in the first year of
Pb+Pb collisions, we conclude that there will be few continuum contributions 
above 40 GeV/$c^2$, evident from the high mass dimuon
distribution published by the CMS \cite{zboson}, in agreement with the result
shown in Fig.~\ref{DiLepCMST}(d). The second Pb+Pb run in 2011 has 20 times more
events which will help quantify the heavy flavour contribution after 
uncorrelated pairs are eliminated by background subtraction techniques.
 Their yields relative to $pp$ collisions at the 
same energy can be used as a high 
statistics probe of the medium properties in Pb+Pb colliions.

\section{Acknowledgments}
The authors are grateful to Dr. D. K. Srivastava for many fruitful discussions.
The work of R.V.\ was performed under the auspices of the U.S.\ Department
of Energy by Lawrence Livermore National Laboratory under Contract
DE-AC52-07NA27344 and also supported by the JET Collaboration.

\end{document}